\documentclass[12pt]{article}
\usepackage[dvips]{color}
\usepackage{epsfig}
\usepackage{amsmath}
\usepackage{graphicx}
\def\Box{\hbox{$\rlap{$\sqcup$}\sqcap$}}
\begin{document}

\setcounter{page}{1}

\pagestyle{plain} \vspace{1cm}
\begin{center}
\Large{\bf Non-Minimal Warm Inflation and Perturbations
 on the Warped DGP Brane with Modified Induced Gravity}\\
\small \vspace{1cm} {\bf Kourosh Nozari$^{a,b,}$
\footnote{knozari@umz.ac.ir}}, \quad {\bf M. Shoukrani$^{a,}$
\footnote{m.shoukrani@umz.ac.ir}}\quad and\quad {\bf  B.
Fazlpour$^{a,}$ \footnote{ b.fazlpour@umz.ac.ir}}\\
\vspace{0.5cm} {\it $^{a}$Department of Physics, Faculty of Basic
Sciences,\\
University of Mazandaran,\\
P. O. Box 47416-95447, Babolsar, IRAN}\\
\vspace{0.5cm} $^{b}${\it Research Institute for Astronomy and
Astrophysics of Maragha,\\ P. O. Box 55134-441, Maragha, IRAN}

\end{center}
\vspace{1.5cm}
\begin{abstract}
We construct a warm inflation model with inflaton field
non-minimally coupled to induced gravity on a warped DGP brane. We
incorporate possible modification of the induced gravity on the
brane in the spirit of $f(R)$-gravity. We study cosmological
perturbations in this setup. In the case of two field inflation such
as warm inflation, usually entropy perturbations are generated.
While it is expected that in the case of one field inflation these
perturbations to be removed, we show that even in the absence of the
radiation field, entropy perturbations are generated in our setup
due to non-minimal coupling and modification of the induced gravity.
We study the effect of dissipation on the inflation parameters of
this extended braneworld scenario.
\end{abstract} \vspace{2cm}
\newpage

\section{Introduction}
The idea of inflation is a very successful paradigm to solve the
problems of the standard cosmology and it provides a basis for
production and evolution of seeds for large scale structure of the
universe [1,2]. From a thermodynamical viewpoint, there are two
possible alternatives to inflationary dynamics: Standard picture is
isentropic inflation referred to as supercooled inflation. In this
picture, universe expands in inflation phase and its temperature
decrease rapidly. When inflation ends, a reheating period introduces
radiation into the universe. The fluctuations in this type of
inflation model are zero-point ground state fluctuations and
evolution of the inflaton field is governed by ground state
evolution equations. In this model we have not any thermal
perturbations and therefore density perturbations are adiabatic ( or
curvature). The other picture is a non-isentropic inflation, the so
called warm inflation. Warm inflation has no need to introduce
reheating phase since interaction between the inflaton and other
fields in this scenario produces the radiation energy density. In
this picture, inflation terminates smoothly and radiation regime is
dominated without a reheating period. The fluctuations during warm
inflation emerge from some excited states and the evolution of the
inflaton has dissipative terms arising from interaction of the
inflaton and other fields [3,4] which affects the inflaton dynamics
through a noise term in the equations of motion [5,6] ( for a
comprehensive list of the references on warm inflation, see [7]) .
The important point in the warm inflation scenario is that the
density fluctuations in this scenario arise from thermal rather than
vacuum fluctuations [3,6,8] and the fluctuations in the radiation
produce the entropy (isocurvature) perturbations. The thermal
fluctuations during warm inflation lead to production of necessary
initial seeds for Large Scale Structure (LSS) formation. The warm
inflation ends when the radiation is dominated in the universe and
the universe enters in a standard Big Bang phase [4,9]. The entropy
fluctuations disappear before
inflation ends.\\
The goal of this investigation is to study cosmological
perturbations in a braneworld viewpoint of the warm inflation in the
presence of interaction between inflaton and modified induced
gravity on the brane. Among various braneworld scenarios, the model
proposed by Dvali, Gabadadze and Porrati (DGP) [10] predicts
deviations from the standard $4$-dimensional gravity even over large
distances. In this scenario, the transition between four and
$5$-dimensional gravitational potentials arises due to the presence
of an induced gravity term in the brane action. Existence of a
higher dimensional embedding space allows for the existence of bulk
or brane matter which can certainly influence the cosmological
dynamics on the brane. In the DGP setup, the bulk is a flat
Minkowski spacetime, but a reduced gravity term appears on the brane
without tension. This model has a rich phenomenology discussed in
[11]. Maeda, Mizuno and Torii have constructed a braneworld scenario
which combines the Randall-Sundrum II ( RS II) and the DGP model
[12]. In this combination, an induced curvature term appears on the
brane in the RS II scenario which contains an AdS bulk. This model
has been called the warped DGP braneworld in the literature [13].
The supercooled inflation models in this scenario were studied in
minimal and non-minimal cases [13-17]. Warm inflation model on a
warped DGP brane in the minimal case has been studied by del Campo
and Herrera [18], but here we consider the effects of the
non-minimal coupling of the scalar field and modified induced
gravity on the brane because one important feature of the
inflationary paradigm is the fact that inflaton can interact with
other fields such as gravitational sector of the theory. This
interaction is shown by the non-minimal coupling of the inflaton
field and modified induced curvature in the spirit of the
scalar-tensor theories, which is motivated from several compelling
reasons ( for a discussion on the reasons to include an explicit
non-minimal coupling between inflaton and gravitational sector in a
typical inflation model, see [19]). In fact, inclusion of the
non-minimal coupling in our setup is not just a matter of taste; it
is forced upon us since as has been indicated in [19], in most
theories used to describe inflationary scenarios, it turns out that
a non-vanishing value of the coupling constant
cannot be avoided.\\
To have a complete treatment of the problem, we consider possible
modification of the induced gravity on the brane in the spirit of
the $f(R)$-gravity. The main motivation for adopting such a
framework is the fact that although inflation is an elegant scenario
to resolve some shortcomings of the standard cosmology and provides
a causal and predictive theory of structure formation, there are
some important and yet unsolved problems in it [20]. Hierarchy
problem, trans-Planckian problem and singularity problem are among
these problems. Modification of the induced gravity in the spirit of
$f(R)$ theories may provide a reliable framework to solve at least
part of these problems. In another words, incorporation of the
modified induced gravity on the brane in the spirit of higher order
gravitational theories may shed more light on these problems. In
fact, $f(R)$-gravity is the simplest way to achieve this goal and it
is also possible to obtain a nonsingular cosmology in this setup.
For a review on $f(R)$-gravity and also inflation and cosmic
acceleration in modified gravity see [21]. In this paper we consider
the general form of $f(R)$-gravity and we discuss cosmological
perturbations in the framework of non-minimal warm inflation on the
warped DGP brane.\\
Through this paper, a dot on a quantity represents the time
derivative and a prim marks derivative with respect to Ricci
curvature $R$.

\section{Warped DGP Scenario}
Consider a 5-dimensional AdS bulk spacetime with a single
$4$-dimensional brane embedded in it. Standard matter, including
inflaton field, are localized on the brane but gravity and possibly
non-standard matter are free to propagate in the bulk. The gravity
is induced on the brane via interaction of bulk gravitons and matter
localized on the brane. The action of this extension of DGP scenario
where brane is foliated in the bulk with warped geometry can be
written as follows [12]
\begin{equation}
{\cal{S}}=\int_{bulk}d^{5}X\sqrt{-{}^{(5)}g}\bigg[\frac{1}{2\kappa_{5}^{2}}
{}^{(5)}R+{}^{(5)}{\cal{L}}_{m}\bigg]+\int_{brane}d^{4}x\sqrt{-g}\bigg[\frac{1}{\kappa_{5}^{2}}
K^{\pm}+{\cal{L}}_{brane}(g_{\alpha\beta},\psi)\bigg].
\end{equation}
In this action the quantities are defined as follows: $X^{A}$ with
$A=0,1,2,3,5$ are coordinates in the bulk while $x^{\mu}$ with
$\mu=0,1,2,3$ are induced coordinates on the brane. $\kappa_{5}^{2}$
is the 5-dimensional gravitational constant. ${}^{(5)}R$ and
${}^{(5)}{\cal{L}}_{m}$ are 5-dimensional Ricci scalar and matter
Lagrangian respectively. $K^{\pm}$ is the trace of the extrinsic
curvature on either sides of the brane. This term is known as the
York-Gibbons-Hawking term [22] which provides a through framework
for imposing suitable boundary conditions on the field equations.
${\cal{L}}_{brane}(g_{\alpha\beta},\psi)$ is the effective
$4$-dimensional Lagrangian. This action is actually a combination of
the Randall-Sundrum II [23] and the DGP model [10]. Consider the
brane Lagrangian as follows
\begin{equation}
{\cal{L}}_{brane}(g_{\alpha\beta},\psi)=\frac{\mu^2}{2}R-\lambda+L_{m}
\end{equation}
where $\mu$ is a mass parameter, $R$ is Ricci scalar of the brane,
$\lambda$ is tension of the brane and $L_{m}$ is Lagrangian of the
matters localized on the brane. Assume that bulk contains only a
cosmological constant, $^{(5)}\Lambda$. With these choices, action
(1) gives either a generalized DGP or a generalized RS II model: it
gives DGP model if $\lambda=0$ and $^{(5)}\Lambda=0$ and gives RS II
model if $\mu=0$ [12].

Considering a spatially flat FRW metric on the brane, the
cosmological dynamics on the brane is given by
\begin{equation}
H^{2}=\frac{1}{3\mu^2}\bigg[\rho+\rho_{0}\Big(1+\varepsilon
{\cal{A}}(\rho,a)\Big)\bigg],
\end{equation}
where $\varepsilon=\pm 1$ is corresponding to two possible branches
of the solutions in this warped DGP model as a manifestation of the
two possible embedding of brane in the bulk. Other quantities are
defined as \,
${\cal{A}}=\bigg[{\cal{A}}_{0}^{2}+\frac{2\eta}{\rho_{0}}\Big
(\rho-\mu^{2}\frac{{\cal{E}}_{0}}{a^{4}}\Big)\bigg]^{1/2}$ where
\,\, ${\cal{A}}_{0}\equiv
\bigg[1-2\eta\frac{\mu^{2}\Lambda}{\rho_{0}}\bigg]^{1/2}$,\,\, $\eta
\equiv\frac{6m_{5}^{6}}{\rho_{0}\mu^{2}}$\,\, with $0<\eta\leq1$
\,\,and \,\,$\rho_{0}\equiv
m_{\lambda}^{4}+6\frac{m_{5}^{6}}{\mu^{2}}$. Note that by
definition, $m_{\lambda}= \lambda^{1/4}$,\, $m_{5}=k_{5}^{-2/3}$\,\,
and ${\cal{E}}_{0}$ is an integration constant where corresponding
term in the generalized Friedmann equation is called the dark
radiation term. Since we are interested in the inflation dynamics of
our model, we neglect dark radiation term in which follows\footnote{
Note however that dark radiation in the background which is
constraint by observations to be a small fraction of the radiation
energy density, has interesting effects in the radiation era. As has
been shown in Ref. [24], on large scales this term slightly
suppresses the radiation density perturbations at late times. In a
kinetic era, this suppression is much stronger and drives the
density perturbations to zero.}. In this case, generalized Friedmann
equation (3) attains the following form
\begin{equation}
H^{2}=\frac{1}{3\mu^2}\bigg[\rho+\rho_{0}+\varepsilon\rho_{0}\Big
({\cal{A}}_{0}^{2}+\frac{2\eta\rho}{\rho_{0}}\Big)^{1/2}\bigg].
\end{equation}
This equation is the basis of our forthcoming arguments.

\section{Non-Minimal $f(R)$-DGP-Inspired Warm Inflation}
After introducing the warped DGP braneworld scenario, here we
consider the case of non-minimal warm inflation in this setup. We
assume that the warm inflation is driven by the non-minimally
coupled scalar field $\varphi$ with potential $V(\varphi)$ on the
warped DGP brane, where possible modification of the induced gravity
on the brane is taken into account within the general framework of
$f(R)$-gravity. The action of this model with a non-minimally
coupled scalar field is given as follows
$${\cal{S}}=\int_{bulk}d^{5}X\sqrt{-{}^{(5)}g}\bigg[\frac{1}{2\kappa_{5}^{2}}
{}^{(5)}R+{}^{(5)}{\cal{L}}_{m}\bigg]$$
\begin{equation}
+\int_{brane}d^{4}x\sqrt{-g}\bigg[\frac{\mu^2}{2}R+\frac{1}{2}\xi
f(R)\varphi^{2}-\frac{1}{2}\partial_{\mu}\varphi\partial^{\mu}\varphi-V(\varphi)+\frac{1}{\kappa_{5}^{2}}
K^{\pm}-\lambda+L'_{m}\bigg].
\end{equation}
where $\xi $ is a non-minimal coupling and $f(R)$ is a function of
the Ricci scalar on the brane [25]. $L'_{m}$ is Lagrangian of the
other matters localized on the brane.

Variation of the action with respect to $\varphi$ gives the equation
of motion of the scalar field in this warm inflation scenario
\begin{equation}
\ddot{\varphi}+3H\dot{\varphi}-\xi f(R)\varphi
+\frac{dV}{d\varphi}=-\Gamma \dot{\varphi}.
\end{equation}
$\Gamma$ is the dissipation coefficient and during inflation period,
it is responsible for decay of the scalar field into radiation.
There are several choices for $\Gamma$, that is: a constant, a
function of the scalar field $\varphi$, a function of temperature
$T$ and a function of both scalar field and temperature,
$(\varphi,T)$ ( for a recent progress in this direction, see [26]).
In the supercooled inflation models $\Gamma=0$ and the equation of
motion has the standard form $\ddot{\varphi}+3H\dot{\varphi}-\xi
f(R)\varphi +\frac{dV}{d\varphi}=0$. The energy-momentum tensor of a
scalar field non-minimally coupled to induced gravity for a
DGP-inspired $f(R)$-gravity scenario is
$$T_{\mu\nu}=g_{\mu \nu}\Big(\frac{1}{2}\xi
 f(R)\varphi^2-\frac{1}{2}g^{\alpha\beta}
 \partial_{\alpha}\varphi\partial_{\beta}\varphi-V(\varphi)\Big)$$
\begin{equation}
+\partial_\mu\varphi\partial_\nu\varphi-\xi f'(R)R_{\mu
\nu}\varphi^{2}
 -\xi\Big(g_{\mu
 \nu}\Box-\nabla_{\mu}\nabla_{\nu}\Big)f'(R)\varphi^{2}.
\end{equation}
So, the energy density and pressure are given by
$$\rho_{\varphi}=\frac{1}{2}\dot{\varphi}^{2}+V(\varphi)+\rho^{(curve)}\hspace{12.5cm}$$
\begin{equation}
=\frac{1}{2}\dot{\varphi}^{2}+V(\varphi)+\xi\Bigg[-\frac{1}{2}
f(R)\varphi^{2}-6 f'(R) \varphi H \dot{\varphi}+3
f'(R)\varphi^{2}(\dot{H}+H^{2})-18 f''(R)\varphi^{2}
H(\ddot{H}+4H\dot{H})\Bigg],\hspace{1cm}
\end{equation}
$$p_{\varphi}=\frac{1}{2}\dot{\varphi}^{2}-V(\varphi)+P^{(curve)}\hspace{12.5cm}$$
$$=\frac{1}{2}\dot{\varphi}^{2}-V(\varphi)+\xi\Bigg[2\big(
\varphi\ddot{\varphi}+2  \varphi H \dot{\varphi}+
\dot{\varphi}^{2}\big)f'(R)+\Big(\frac{1}{2}f(R)-f'(R)(\dot{H}+3H^{2})\Big)\varphi^{2}\hspace{2cm}$$
\begin{equation}
+12f''(R)\Big(H\varphi^{2}+2\varphi\dot{\varphi}\Big)(\ddot{H}+4H\dot{H})
+6 f''(R)(\dddot {H}+4\dot{H}^{2}+4H\dot{H})\varphi^{2}+36
f'''(R)(\ddot{H}+4H\dot{H})^{2}\varphi^{2}\Bigg].
\end{equation}
Note that $\rho^{(curve)}$ and $p^{(curve)}$ are curvature-dependent
parts of the energy density and pressure of the non-minimally
coupled scalar field respectively. The conservation equation for
scalar field energy density in this dissipative setup is given by
\begin{equation}
\dot{\rho}_\varphi+3H(\rho_\varphi+P_\varphi)=-\Gamma
\dot{\varphi}^2.
\end{equation}
Since $\Gamma$ is responsible for interaction of the scalar field
and radiation, it is expected that this coefficient has no
dependence on the non-minimal coupling of the scalar field and
$f(R)$-gravity. The energy density and pressure that contain both
scalar field and radiation contributions are
\begin{equation}
\rho=\rho_{\varphi}+\rho_{\gamma}=\rho_{\varphi}+\frac{3}{4}ST,
\end{equation}
and
\begin{equation}
p=p_{\varphi}+p_{\gamma}=p_{\varphi}+\frac{1}{4}ST,
\end{equation}
where $\rho_{\gamma}$ and $p_{\gamma}$ are the radiation energy
density and pressure respectively. The conservation equation for the
combined system of scalar field and radiation is given by
 \begin{equation}
\dot{\rho}+3H(\rho+P)=0,
\end{equation}
which implies the entropy production. Making use of Eq. (11) and
(12) we get
 \begin{equation}
T(\dot{S}+3HS)=\Gamma\dot{\varphi}^{2},
\end{equation}
where by equation (6),\, $\dot{\varphi}$ is directly related to the
non-minimal coupling. The basic idea of warm inflation is that
radiation production is occurring concurrently with inflationary
expansion due to dissipation of the inflaton field system. The
equation of state for radiation field is given by
$P_{\gamma}=\frac{\rho_{\gamma }}{3}$. Therefore, the conservation
equation of $\rho_\gamma$ yields the following result
\begin{equation}
\dot{\rho}_\gamma+4H\rho_\gamma=\Gamma\dot{\varphi}^2.
\end{equation}
that a part of dynamics is represented by this equation.\\
In the slow-roll approximation where $\ddot{\varphi}\ll V(\varphi)$,
equation of motion for the scalar field takes the following form
\begin{equation}
\dot{\varphi}=\frac{\xi f(R)\varphi-V_{,\varphi}}{\Gamma+3H},
\end{equation}
where $ V_{,\varphi}\equiv \frac{dV}{d\varphi}$. In warm inflation,
the radiation production is quasi-stable so that \,
$\dot{\rho}_{\gamma}\ll 4H\rho_{\gamma}$ \, and \,
$\dot{\rho}_{\gamma}\ll\Gamma\dot{\varphi}^{2}$. Therefore we have
from (15)
\begin{equation}
\rho_{\gamma}=\frac{\Gamma\dot{\varphi}^{2}}{4H}.
\end{equation}
By using equations (4), (16) and (17) we obtain
\begin{equation}
\rho_{\gamma}=\alpha
T^{4}=\frac{r\mu^{2}}{4(1+r)^{2}}\Bigg[\frac{\Big(\xi
f(R)\varphi-V_{,\varphi}\Big)^{2}}{\Big[\rho_{\varphi}+\rho_{0}+
\varepsilon\rho_{0}\Big({\cal{A}}_{0}^{2}+\frac{2\eta\rho_{\varphi}}{\rho_{0}}\Big)^{1/2}\Big]}\Bigg],
\end{equation}
where we have used the definition of the dissipation factor as
follows
\begin{equation}
r\equiv\frac{\Gamma}{3H},
\end{equation}
which is a dimensionless parameter. Since during an inflationary era
the scalar field energy density dominates over the energy density of
the radiation field, that is,\, $\rho_{\varphi}>\rho_{\gamma}$,\, we
can assume $\rho\simeq\rho_{\varphi}$. Here $\alpha\equiv
\frac{g_{*}\pi^{2}}{30}$ is the Stefan-Boltzmann constant and
$g_{*}$ is the number of degrees of freedom for the radiation field,
that in the standard cosmology is $g_{*}\approx 100$. A part of the
effects of the non-minimal coupling and dissipation is hidden in the
definition of energy density, $\rho_{\varphi}$, which attains the
following form by using (8)
\begin{equation}
\rho_{\varphi}\approx V+\xi\Bigg[-\frac{1}{2} f(R)\varphi^{2}-2
\frac{f'(R)}{(1+r)} \Big(\xi
f(R)\varphi^{2}-V_{,\varphi}\varphi\Big)+\frac{1}{2}
f'(R)R\varphi^{2}+ 3 f'(R)\varphi^{2}H^{2}-3f''(R)\dot{R} H\Bigg],
\end{equation}
where $R=6(\dot{H}+2H^{2})$ and $\dot{R}=6(\ddot{H}+4H\dot{H})$.\\
We define the most important slow-roll parameter as\footnote{Note
that we use $\epsilon$ for slow-roll parameter while
$\varepsilon=\pm 1$ marks two possible branches of the DGP setup.}
\begin{equation}
\epsilon\equiv
-\frac{\dot{H}}{H^{2}}=\frac{\mu^{2}}{2(1+r)}\big(V_{,\varphi}^{2}+\xi\beta
\big)\frac{\bigg[1+
\varepsilon\eta\Big({\cal{A}}_{0}^{2}+\frac{2\eta\rho_{\varphi}}{\rho_{0}}\Big)^{-1/2}\bigg]
}{\bigg[\rho_{\varphi}+\rho_{0}+\varepsilon\rho_{0}\Big({\cal{A}}_{0}^{2}+\frac{2\eta\rho_{\varphi}}
{\rho_{0}}\Big)^{1/2}\bigg]^{2}},\hspace{.7cm}
\end{equation}
where by definition
$$\beta\equiv V_{,\varphi}\Bigg(-2f(R)\varphi+f'(R)\Big[-2\frac{V_{,\varphi}\,r_{,\varphi}
\,\varphi}{(1+r)^{2}}+2\frac{V_{,\varphi\varphi}\,\varphi}{(1+r)}
-2\frac{V_{,\varphi}}{(1+r)}+(R+6H^{2})\varphi+\frac{6\varphi^{2}H\dot{H}}{\dot{\varphi}}\Big]$$
\begin{equation}
+\frac{f''(R)}{\dot{\varphi
}}\Big[2\frac{V_{,\varphi}\,\varphi\dot{R}}{(1+r)}+\frac{1}{2}(R+6H^{2})\dot{R}\varphi^{2}
-3(\dot{R}\dot{H}+\ddot{R}H
)\Big]-3f'''(R)\frac{\dot{R}^{2}H}{\dot{\varphi}}\Bigg).
\end{equation}
Due to complicated form of the equations, here we restrict our
analysis to the first order of the non-minimal coupling\footnote{
Note that this assumption is justified since $\xi$ is constraint to
be very close to the conformal coupling, $\xi=\frac{1}{6}$ by the
recent observations ( see [27] for instance).}, $\xi$. Since $\beta$
itself is multiplied by $\xi$ in equation (21), to have a first
order analysis we should consider those terms of $\beta$ that are
independent of $\xi$. So, we should consider only the terms
independent of $\xi$ in the definition of $H$ and $\dot{H}$. These
terms are \,$H=-\frac{V_{,\varphi}}{3(1+r)\dot{\varphi}}$ \, and
$\dot{H}=-\frac{V_{,\varphi\varphi}}{3(1+r)}+\frac{V_{,\varphi
}\,r_{,\varphi}}{3(1+r)^{2}}+\frac{V_{,\varphi}\ddot{\varphi}}{3(1+r)\dot{\varphi}^{2}}$,\,
respectively.\\
In comparison with minimal warm inflation on DGP brane as presented
in Ref. [18], we see that our equation (21) reduces to equation (15)
of Ref. [18] for $\xi=0$. On the other hand, for $r=0$ we recover
typical expression for non-minimal supercool inflation in the warped
DGP brane. The relation between energy densities of radiation and
inflaton fields can be calculated using the slow-roll parameter
$\epsilon$ to find
\begin{equation}
\rho_{\gamma}=\frac{r}{2(1+r)}\epsilon\,\frac{\Big(V_{,\varphi}-\xi
f(R)\varphi\Big)^{2
}}{\Big(V_{,\varphi}^{2}+\xi\beta\Big)}\times\Bigg[\frac{\rho_{\varphi}+\rho_{0}
\Big[1+\varepsilon\Big({\cal{A}}_{0}^{2}+\frac{2\eta\rho_{\varphi}}{\rho_{0}}\Big)^{1/2}\Big]}{1+\varepsilon
\eta\Big({\cal{A}}_{0}^{2}+\frac{2\eta\rho_{\varphi}}{\rho_{0}}\Big)^{-1/2}}\Bigg].
\end{equation}
The inflation takes place when the condition $\epsilon<1$ (or
equivalently $\ddot{a}>0$) is fulfilled. This condition in our case
reduces to the following expression for realization of the warm
inflation in our non-minimal setup
\begin{equation}
\Bigg[\frac{\rho_{\varphi}+\rho_{0}
\Big[1+\varepsilon\Big({\cal{A}}_{0}^{2}+\frac{2\eta\rho_{\varphi}}{\rho_{0}}\Big)^{1/2}\Big]}{1+\varepsilon
\eta\Big({\cal{A}}_{0}^{2}+\frac{2\eta\rho_{\varphi}}{\rho_{0}}\Big)^{-1/2}}\Bigg]>
\frac{2(1+r)}{r}\rho_{\gamma}\,\frac{\Big(V_{,\varphi}^{2}+\xi\beta\Big)
}{\Big(V_{,\varphi}-\xi f(R)\varphi\Big)^{2}}.
\end{equation}
The warm inflationary period lasts up to violation of this
condition. The inflationary epoch ends when the $\epsilon\simeq 1$
is fulfilled and this implies that
\begin{equation}
\Bigg[\frac{\rho_{\varphi}+\rho_{0}
\Big[1+\varepsilon\Big({\cal{A}}_{0}^{2}+\frac{2\eta\rho_{\varphi}}{\rho_{0}}\Big)^{1/2}\Big]}{1+\varepsilon
\eta\Big({\cal{A}}_{0}^{2}+\frac{2\eta\rho_{\varphi}}{\rho_{0}}\Big)^{-1/2}}\Bigg]=
\frac{2(1+r)}{r}\rho_{\gamma}\,\frac{\Big(V_{,\varphi}^{2}+\xi\beta\Big)
}{\Big(V_{,\varphi}-\xi f(R)\varphi\Big)^{2}}.
\end{equation}
The second slow-roll parameter in this setup is given by
$$\alpha \equiv -\frac{\ddot{H}}{H\dot{H}}\simeq\mu^{2}\Big[\frac{V_{,\varphi\varphi}}{(1+r)
}-\frac{V_{,\varphi}\,r_{,\varphi}}{(1+r)^{2}}
-\frac{\xi}{(1+r)}\big(f'(R)\dot{R}+f(R)-\frac{f(R)r_{,\varphi}\,\varphi}{(1+r)}\big)\Big]$$
\begin{equation}
\times\bigg[\rho_{\varphi}+\rho_{0}+\varepsilon\rho_{0}\Big({\cal{A}}_{0}^{2}+\frac{2\eta
\rho_{\varphi}}{\rho_{0}}\Big)^{1/2}\bigg]^{-1}.
\end{equation}
In the minimal case one has only the first two terms of the right
hand side of this expression. Note also that in our setup we
consider $\Gamma=\Gamma(\varphi)$ or equivalently $r=r(\varphi)$.
The number of e-folds, $N\equiv\ln\frac{a_{e}}{a_{i}}$ in the
presence of the non-minimal coupling and for a warped DGP-inspired
$f(R)$-gravity can be written as
$$N(\varphi)=-\int_{\varphi_{i}}^{\varphi_{e}}3H^{2}\frac{(1+r)}{\Big(V_{,\varphi}-\xi
f(R)\varphi\Big)}d\varphi\hspace{4.7cm}$$
\begin{equation}
\hspace{2cm}=-\frac{1}{\mu^2}\int_{\varphi_{i}}^{\varphi_{e}}\frac{(1+r)}{\Big(V_{,\varphi}-\xi
f(R)\varphi\Big)}\bigg[\rho_{\varphi}+\rho_{0}+\varepsilon
\rho_{0}\Big({\cal{A}}_{0}^{2}+\frac{2\eta\rho_{\varphi}}{\rho_{0}}\Big)^{1/2}\bigg]d\varphi.
\end{equation}
where $\varphi_{i}$ denotes the value of the scalar field $\varphi$
when Universe scale observed today crosses the Hubble horizon during
inflation, and $\varphi_{e}$ is the value of the scalar field when
the Universe exits the inflationary phase.

\section{Perturbations}
The inhomogeneous perturbations of the FRW background are described
by the metric in the longitudinal gauge [28, 29]
\begin{equation}
ds^{2}=-\big(1+2\phi\big)dt^{2}+a^{2}(t)\big(1-2\psi\big)\delta_{i\,j}\,dx^{i}dx^{j}.
\end{equation}
where $a(t)$ is the scale factor on the brane, $\phi = \phi(t, x)$
and $\psi=\psi(t, x)$ are the metric perturbations. The radiation
and scalar fields interact through the friction term $\Gamma$. The
spatial dependence of all perturbed quantities are of the form of
plane waves $e^{ik.x}$, where $k$ is the wave number. A perturbation
of the metric implies, through Einstein's equations of motion, a
perturbation in the energy-momentum tensor. The energy-momentum
tensor in our setup as defined in equation (7) is diagonal if we
note that $R$ is just a function of the cosmic time [30]. We note
that the two metric perturbations are not equal due to the presence
of the anisotropic stress perturbation. The perturbed Weyl
contribution to the Einstein equations can be parametrized as an
effective fluid with anisotropic stress perturbation, and this
contribution cannot be set to zero.\\
The perturbed field equations can be obtained straightforwardly from
Einstein field equations. In a warped DGP braneworld model, the
Einstein field equations change to effective equations on the brane
given as [12]
\begin{equation}
G_{\mu\nu}=\frac{\Pi_{\mu\nu}}{m_{5}^{6}}-E_{\mu\nu},
\end{equation}
where $m_{5}^{6}=\frac{\rho_{0}\mu^{2}\eta}{6}$ \, and
\begin{equation}
\Pi_{\mu\nu}=-\frac{1}{4}{T}_{\mu\sigma}{T}_{\nu}^{\sigma}+
\frac{1}{12}{T}{T}_{\mu\nu}+
\frac{1}{8}g_{\mu\nu}\Big({T}_{\rho\sigma}{T}^{\rho\sigma}-\frac{1}{3}{T}^{2}\Big),
\end{equation}
and  ${T}_{\mu\nu}$ is the total stress-tensor on the brane. Also we
have
\begin{equation}
E_{\mu\nu}=C_{MRNS}\,\, n^{M}\,\,n^{N}
{g^{R}}_{\mu}\,\,{g^{S}}_{\nu}
\end{equation}
where $C_{MRNS}$ is the five dimensional Weyl tensor and $n_{A}$ is
the spacelike unit vector normal to the brane. The Friedmann
equation (3) can be calculated directly from these equations (see
[31] for instance). So, to obtain perturbed field equations, if we
adopt the standard prescription as has been presented in Ref. [32],
we should replace the quantities in the standard picture with
corresponding effective quantities. We note that in the background
spacetime, $E_{\mu\nu}=0$ and we can use equation (4) as Friedmann
equation in this setup. But the perturbed FRW brane has a nonzero
$E_{\mu\nu}$, which encodes the effects of the bulk gravitational
field on the brane [33] and we have use the Friedmann equation (3)
where $E_{\,0}^{0}=\frac{{\cal{E}}_{0}}{a^{4}}$ [12]. The perturbed
$5D$ field equations are needed to determine the evolution of
$\delta E_{\mu\nu}$. In this manner, the temporal part of the
perturbed field equations are given as
\begin{equation}
-3H(H\phi+\dot{\psi})-\frac{k^{2}}{a^{2}}=\frac{1}{2\mu^{2}}\delta
\rho_{tot}
\end{equation}
\begin{equation}
\ddot{\psi}+3H(H\phi+\dot{\psi})+H\dot{\phi}+2\dot{H}\phi+\frac{1}{3a^{2}}k^{2}(\phi-\psi)=
\frac{1}{2\mu^{2}}\delta P_{tot}
\end{equation}
\begin{equation}
\dot{\psi}+H\phi=\frac{1}{2\mu^{2}}\Big(-\frac{4}{3k}\rho_{\gamma}av+
\frac{\rho_{\varphi}}{\rho_{0}\eta}\dot{\varphi}\delta\varphi
-\frac{\rho_{\varphi}}{\rho_{0}\eta}\int(\delta T_{i}^{0})_{\xi}\,d
x_{i}\Big)+\frac{1}{2}\int(\delta E_{i}^{0})\,d x_{i},
\end{equation}
\begin{equation}
\psi-\phi=\frac{\delta f'(R)}{f'(R)}+8\pi
G\frac{H}{r_{c}(\dot{H}+2H^{2})-H} a^{2}\delta_{\pi E}
\end{equation}
where $v$ appears from the decomposition of the velocity field as
$\delta u_{i} =-\frac{iak_{j}}{k}ve^{ik.x}\,(j=1,2,3)$ [29] and we
have omitted the subscript $k$. The last equation is related to the
perturbed effective Einstein equation for the component $\delta
G_{i}^{0}$ that $\delta\Pi_{i}^{0}=-\frac{1}{6}
\rho_{\varphi}\Big[\partial_{i}\dot{\varphi}\delta\varphi-(\delta
T_{i}^{0})_{\xi}\Big]$. Equations (32) and (33) above have their
standard forms but now in the non-minimal DGP brane world model. The
perturbed total energy density and pressure in longitudinal gauge
can be written as
\begin{equation}
\delta\rho_{tot}=\delta\rho_{eff}+\delta\rho_{\gamma},
\end{equation}
and
\begin{equation}
\delta P_{tot}=\delta P_{eff}+\frac{1}{3}\delta\rho_{\gamma},
\end{equation}
respectively. The second terms on the right hand side of these two
equations are hallmark of the warm inflation, because a perturbation
of the metric leads to a perturbation in the stress energy-momentum
tensor and in the warm inflationary model the stress-momentum tensor
contains the radiation field too. In the DGP brane world model by
using the Friedmann equation (3), one can define an effective
gravitational energy density and pressure as
\begin{equation}
\rho_{eff}=\rho_{\varphi}+\rho_{0}+\varepsilon\rho_{0}\Big[{\cal{A}}_{0}^{2}+\frac{2
\eta}{\rho_{0}}\big(\rho_{\varphi}-\frac{\mu^{2}{\cal{E}}_{0}}{a^{4}}\big)\Big]^{1/2},
\end{equation}
and
$$P_{eff}=P_{\varphi}+\varepsilon\eta\Big(P_{\varphi}+\rho_{\varphi}-\frac{4}{3}\frac{\mu^{2}
{\cal{E}}_{0}}{a^{4}}\Big)\Big[{\cal{A}}_{0}^{2}+\frac{2
\eta}{\rho_{0}}\big(\rho_{\varphi}-\frac{\mu^{2}{\cal{E}}_{0}}{a^{4}}\big)\Big]
^{-1/2}$$
\begin{equation}
-\Bigg(\rho_{0}+\varepsilon\rho_{0}\Big[{\cal{A}}_{0}^{2}+\frac{2
\eta}{\rho_{0}}\big(\rho_{\varphi}-\frac{\mu^{2}{\cal{E}}_{0}}{a^{4}}\big)\Big]^{1/2}\Bigg),
\end{equation}
respectively where  $\rho_{eff}$ and $P_{eff}$ obey the standard
Friedmann equation. In other words, using the standard Friedmann
equation as $H^{2}=\frac{1}{3\mu^{2}}\rho$ and substituting for
$\rho$ from equation (38), we recover the equation (3). The
effective pressure is then calculated by using the continuity
equation. Now we can rewrite equations. (36) and (37) for DGP brane
world model as
\begin{equation}
\delta\rho_{tot}=\delta\rho_{\varphi}+\varepsilon\eta\Big[{\cal{A}}_{0}^{2}+\frac{2
\eta}{\rho_{0}}\big(\rho_{\varphi}-\mu^{2}E_{\,0}^{0}\big)\Big]^{-1/2}
\big(\delta\rho_{\varphi}-\mu^{2}\delta
E_{\,0}^{0}\big)+\delta\rho_{\gamma},
\end{equation}
$$\delta P_{tot}=\delta
P_{\varphi}+\varepsilon\eta\Big[{\cal{A}}_{0}^{2}+\frac{2
\eta}{\rho_{0}}\big(\rho_{\varphi}-\mu^{2}E_{\,0}^{0}\big)\Big]^{-1/2}\Big(\delta
P_{\varphi}-\frac{1}{3}\mu^{2}\delta E_{\,0}^{0}\Big)$$
\begin{equation}
-\frac{\varepsilon\eta^{2}}{\rho_{0
}}\Big(P_{\varphi}+\rho_{\varphi}-\frac{4}{3}\mu^{2}
E_{\,0}^{0}\Big)\Big[{\cal{A}}_{0}^{2}+\frac{2
\eta}{\rho_{0}}\big(\rho_{\varphi}-\mu^{2}E_{\,0}^{0}\big)\Big]
^{-3/2}\big(\delta\rho_{\varphi}-\mu^{2}\delta
E_{\,0}^{0}\big)+\frac{1}{3}\delta\rho_{\gamma},
\end{equation}
where $E_{\,0}^{0}$ can calculated from the general equation $\delta
E_{\,\nu}^{\mu}$ as
\begin{equation}
\delta E_{\,\nu}^{\mu}=-\frac{1}{\mu^{2}}\left(%
\begin{array}{cc}
  -\delta\rho_{E} & a \delta q_{E}\\
  a^{-1} \delta q_{E} & \frac{1}{3}\delta\rho_{E}\delta^{i}_{\,j}+\delta\pi^{i}_{\,E\,j}
  \\
\end{array}%
\right),
\end{equation}
where $E_{\,\nu}^{\mu}$ can parametrize as an effective fluid , with
density perturbation $\delta\rho_{E}$, isotropic pressure
perturbation $\frac{1}{3}\,\delta\rho_{E}$, anisotropic stress
perturbation $\delta\pi_{E}$ and energy flux perturbation $\delta
q_{E}$. Indeed, the perturbed Weyl contribution to the Einstein
equations can be parametrized as an effective fluid with anisotropic
stress perturbation (for details, see [33]).

$\delta\rho_{\varphi}$ and $\delta P_{\varphi}$ contain the effects
of the non-minimal coupling
\begin{equation}
\delta\rho_{\varphi}=\dot{\varphi}\delta\dot{\varphi}-\dot{\varphi}^{2}\phi+
V_{,\varphi}\delta\varphi+\delta\rho_{\xi}\,,
\end{equation}
\begin{equation}
\delta
P_{\varphi}=\dot{\varphi}\delta\dot{\varphi}-\dot{\varphi}^{2}\phi-V_{,\varphi}\delta\varphi+\delta
P_{\xi}\,.
\end{equation}
The last terms in both of these relations are related to the
non-minimal coupling of the scalar filed and induced gravity on the
brane and can be calculated as follows
\begin{equation}
(\delta T_{0}^{0})_{\xi}=(\delta\rho)_{\xi},
\end{equation}
\begin{equation}
(\delta T_{i}^{j})_{\xi}=(\delta P)_{\xi}\,\delta_{i}^{j},
\end{equation}
where
\begin{equation}
(\delta T_{\nu}^{\mu})_{\xi}=\delta g^{\mu \alpha}(T_{\alpha
\nu})_{\xi}+g^{\mu \alpha}(\delta T_{\alpha \nu})_{\xi},
\end{equation}
so that $(T_{\mu\nu})_{\xi}$ and $(\delta T_{\mu\nu})_{\xi}$ are
defined as follows
\begin{equation}
(T_{\mu\nu})_{\xi}=-\xi\varphi^{2}f'(R)R_{\mu\nu}-\xi\Big(g_{\mu
\nu}\Box-\nabla_
{\mu}\nabla_{\nu}\Big)f'(R)\varphi^{2}+\frac{1}{2}\xi g_{\mu
\nu}\varphi^{2}f(R),
\end{equation}
and
$$(\delta T_{\mu\nu})_{\xi}=-\xi\Bigg[\varphi\delta\varphi\Big(2f'(R)R_{\mu\nu}-g_{\mu\nu}f(R)\Big)+\varphi^{2}\delta R
\Big(f''(R) R_{\mu \nu}-\frac{1}{2}g_{\mu
\nu}f'(R)\Big)+\varphi^{2}f'(R)\delta R_{\mu \nu}$$
\begin{equation}
+\Big(g_{\mu \nu}\Box-\nabla_
{\mu}\nabla_{\nu}\Big)\Big(f''(R)\varphi^{2}\delta R+2\varphi
f'(R)\delta\varphi\Big) +\delta g_{\mu
\nu}\Big[\Box(f'(R)\varphi^{2})-\frac{1}{2}\varphi^{2}f(R)\Big]\Bigg].
\end{equation}
We need the following relation to calculate equation (47) explicitly
\begin{equation}
\delta
R=2\Big[(\frac{k^{2}}{a^{2}}-3\dot{H})\phi-\frac{2k^{2}}{a^{2}}\psi-
3\big(\ddot{\psi}+4H\dot{\psi}+H\dot{\phi}+\dot{H}\phi+4H^{2}\phi
\big)\Big]
\end{equation}
To have a complete set of equations for treating perturbations, we
perturb equations (6) and (15) to find
$$\delta
\ddot{\varphi}+(3H+\Gamma)\delta\dot{\varphi}+\Big(V_{,\varphi\varphi}+\frac{k^{2}}{a^{2}}+
\Gamma_{,\varphi}\dot{\varphi}-\xi f(R)\Big)\delta\varphi$$
$$
=\dot{\varphi}(3\dot{\psi}+\dot{\phi})+\phi \Big(2\xi
f(R)\varphi-2V_{,\varphi}+\Gamma\dot{\varphi}\Big)
$$
\begin{equation}
+2\xi
f'(R)\varphi\Big[(\frac{k^{2}}{a^{2}}-3\dot{H})\phi-\frac{2k^{2}}{a^{2}}\psi-
3\big(\ddot{\psi}+4H\dot{\psi}+H\dot{\phi}+\dot{H}\phi+4H^{2}\phi
\big)\Big]
\end{equation}
\begin{equation}
\delta
\dot{\rho}_{\gamma}+4H\delta\rho_{\gamma}+\frac{4}{3}ka\rho_{\gamma}\nu=4\rho_{\gamma}\dot{\psi}
+\dot{\varphi}^{2}\Gamma_{,\varphi}\delta\varphi+\Gamma\dot{\varphi}(2\delta\dot{\varphi}-3\dot{\varphi}\phi)
\end{equation}
We study the effects of the non-minimal coupling of the scalar field
and modified induced gravity on the brane in the warm inflation and
we compare our results with the minimal case. Note that equation
(52) is the same the corresponding equation for minimal case but
other equations mentioned above are changed considerably.

\section{Isocurvature Perturbations}
To interpret the evolution of the cosmological perturbations, the
scalar perturbations can be decomposed so that: a) The projection
orthogonal to the trajectory which is called entropy or isocurvature
perturbation is generated if inflation is driven by more than one
scalar field, and  b) The parallel projection corresponds to the
adiabatic or curvature perturbations and this type of perturbations
are generated if the inflaton field is the only field in inflation
period [34,35]. Note however that these perturbations might even be
cross-correlated to the entropy ones [36-38].\\
Since warm inflation paradigm includes two interacting fields,
isocurvature (entropy) perturbations are expected to be generated
due to thermal fluctuations in the radiation field since the scalar
and radiation fields interact in a thermal bath [32,39,40].\\
For treating entropy perturbations, we note that $\delta P$ and
$\delta\rho$ are related together via entropy perturbation $\delta
S$ [32,41]
\begin{equation}
\big(\dot{P}\delta S=\delta P-c_{s}^{2}\delta\rho\big)_{tot},
\end{equation}
where
$c_{s}^{2}=\frac{\dot{P_{\gamma}}+\dot{P}_{eff}}{\dot{\rho_{\gamma}}+\dot{\rho}_{eff}}$
is the sound effective velocity in the fluid composed of the
radiation and scalar field non-minimally coupled to modified induced
gravity on the warped DGP brane. $\dot{P}\delta S$ is the
non-adiabatic pressure perturbation, $(\dot{P}\delta
S)_{tot}\equiv\delta P_{nad}$, which is due to variation of the
total equation of state that relates $P$ and $\rho$. The entropy
perturbation $\delta S$ represents the displacement between
hypersurfaces of uniform pressure and density.\\
Using equations (40)-(44) in equation (53), we have
$$\delta P_{nad}=(1-c_{s}^{2})\delta\rho_{tot
}-\big(2V_{,\varphi}\delta\varphi+\delta \rho_{\xi}-\delta
P_{\xi}\big)\bigg(1+\varepsilon\eta\Big[{\cal{A}}_{0}^{2}+\frac{2
\eta}{\rho_{0}}\big(\rho_{\varphi}-\mu^{2}E_{\,0}^{0}\big)\Big]^{-1/2}\bigg)$$
$$-\frac{2}{3}\delta\rho_{\gamma}+\frac{2}{3}\varepsilon\eta\mu^{2} \delta
E_{\,0}^{0}\Big[{\cal{A}}_{0}^{2}+\frac{2
\eta}{\rho_{0}}\big(\rho_{\varphi}-\mu^{2}E_{\,0}^{0}\big)\Big]
^{-1/2}$$
\begin{equation}
-\frac{\varepsilon\eta^{2}}{\rho_{0}}\Big(\rho_{\varphi}+P_{\varphi}-\frac{4}{3}
\mu^{2}E_{\,0}^{0}\Big)\Big[{\cal{A}}_{0}^{2}+\frac{2
\eta}{\rho_{0}}\big(\rho_{\varphi}-\mu^{2}E_{\,0}^{0}\big)\Big]
^{-3/2}\big(\delta\rho_{\varphi}-\mu^{2}\delta E_{\,0}^{0}\big).
\end{equation}
Note that if we set $\varepsilon=0$, this expression reduces to the
standard model result and all traces of the DGP setup will
disappear.\,  $\rho_{\varphi}$, $P_{\varphi}$ and
$\delta\rho_{\varphi}$ have been defined by (8), (9) and (43). Using
the equations (32), (33) and (34) we can rewrite this relation as
$$\delta P_{nad}=-\frac{2\mu^{2}(1-c_{s}^{2}-\cal{Y})}{ a^{2}}k^{2}\psi-2\mu^{2}(H\phi+\dot{\psi})\chi
-\frac{2}{3}\delta\rho_{\gamma}+\big(\delta\rho_{\gamma}+\mu^{2}\delta
E_{\,0}^{0}\big){\cal{Y}}+\frac{2}{3}\mu^{2}\delta
E_{\,0}^{0}\big({\cal{Z}}-1\big)$$
\begin{equation}
+\Big(\delta
P_{\xi}-\delta\rho_{\xi}-\frac{8\rho_{\gamma}\rho_{0}aV_{,\varphi}
v\eta}{3k\rho_{\varphi}\dot{\varphi}}-2\frac{V_{,\varphi}}{\dot{\varphi}}
\int(\delta T_{i}^{0})_{\xi}\,dx_{i}+\frac{\rho_{0}V_{,\varphi}
\eta}{\rho_{\varphi}\dot{\varphi}}\int\delta E_{\,i}^{0}\,dx_{i}
\Big){\cal{Z}}.
\end{equation}
where $\chi$, $\cal{Y}$ and ${\cal{Z}}$ in the non-minimal case are
defined as
$$\chi\equiv\frac{8H\rho_{\gamma}-2\Gamma\dot{\varphi}^{2}
-3(2V_{,\varphi}\dot{\varphi}+\dot{\rho}_{\xi}-\dot{P}_{\xi}){\cal{Z}}-8\mu^{2}E_{\,0}^{0}
H\big({\cal{Z}}-1\big)-3{\cal{Y}}
{\cal{Z}}\big(\dot{\varphi}\ddot{\varphi}+V_{,\varphi}\dot{\varphi}+\dot{\rho}_{\xi}+4\mu^{2}E_{\,0}^{0}H\big)
}{3\big(\frac{4}{3}\rho_{\gamma}+\dot{\varphi}^{2}\big)-\frac{1}{H}\big(\xi
f(R)\varphi \dot{\varphi}+\dot{\rho_{\xi}}\big)
{\cal{Z}}+\Big[\big(3+\frac{\Gamma}{H}\big)\dot{\varphi}^{2}-4\mu^{2}E_{\,0}^{0}\Big]\big({\cal{Z}}-1\big)}$$
\begin{equation}
+\frac{2V_{,\varphi}\rho_{0}\eta}{\rho_{\varphi}\dot{\varphi}}{\cal{Z}}-3H{\cal{Y}},
\end{equation}
\begin{equation}
{\cal{Y}}\equiv
\frac{\varepsilon\eta^{2}}{\rho_{0}}\big(P_{\xi}+\rho_{\xi}+\dot{\varphi}^{2}
-\frac{4}{3}\mu^{2}E_{\,0}^{0}\big)
\bigg(\Big[{\cal{A}}_{0}^{2}+\frac{2
\eta}{\rho_{0}}\big(\rho_{\varphi}-\mu^{2}E_{\,0}^{0}\big)\Big]^{3/2}{\cal{Z}}\bigg)^{-1},
\end{equation}
and
\begin{equation}
{\cal{Z}}\equiv\bigg(1+\varepsilon\eta
\Big[{\cal{A}}_{0}^{2}+\frac{2
\eta}{\rho_{0}}\big(\rho_{\varphi}-\mu^{2}E_{\,0}^{0}\big)\Big]^{-1/2}\bigg),
\end{equation}
respectively. We use the slow-roll approximation and quasi-stable
conditions (16) and (17) to write
$$\chi=-2\Gamma+\frac{\Bigg[\dot{\rho_{\xi}}\Big(-\frac{V_{,\varphi}}{\dot{\varphi}}+\Gamma+
\frac{\xi
f(R)\varphi}{\dot{\varphi}}\Big)+\dot{P_{\xi}}\Big(\frac{V_{,\varphi}}{\dot{\varphi}}+\Gamma-
\frac{\xi f(R)\varphi}{\dot{\varphi}}\Big)+2 V_{,\varphi}\big(\xi
f(R)\varphi-V_{,\varphi}\big)
\Bigg]{\cal{Z}}}{\big(V_{,\varphi}\dot{\varphi}+\dot{\rho_{\xi}}\big){\cal{Z}}
+4\mu^{2}E_{\,0}^{0}H\big({\cal{Z}}-1\big)}$$
\begin{equation}
+2\frac{V_{,\varphi}\rho_{0}\eta}{\rho_{\varphi}\dot{\varphi
}}{\cal{Z}}-\frac{4\mu^{2}E_{\,0}^{0}\Big(4\big({\cal{Z}}-1\big)-3{\cal{Y}}\Big)H^{2}
}{\big(V_{,\varphi}\dot{\varphi}+\dot{\rho_{\xi}}\big){\cal{Z}}
+4\mu^{2}E_{\,0}^{0}H\big({\cal{Z}}-1\big)}.
\end{equation}
As is obvious from equation (55), in addition to dissipation, the
non-minimal coupling of the scalar field and modified induced
gravity on the brane has a crucial role in the shape of the entropy
perturbations; it is seen in the first two terms (where a part of
the effects of the non-minimal coupling is hidden in the definition
of the sound effective velocity, $c_{s}^{2}$) and in the last three
terms of this equation. In the minimal case, equation (59) leads to
$$\chi=-2\Gamma-\frac{2V_{,\varphi}^{2}{\cal{Z}}}{V_{,\varphi}\dot{\varphi}{\cal{Z}}
+4\mu^{2}E_{\,0}^{0}H\big({\cal{Z}}-1\big)}+2\frac{V_{,\varphi}\rho_{0}\eta}{\rho_{\varphi}\dot{\varphi
}}{\cal{Z}}-\frac{4\mu^{2}E_{\,0}^{0}\Big(4\big({\cal{Z}}-1\big)-3{\cal{Y}}\Big)H^{2}
}{V_{,\varphi}\dot{\varphi}{\cal{Z}}
+4\mu^{2}E_{\,0}^{0}H\big({\cal{Z}}-1\big)}$$ which contains
dissipation effect in the DGP model. However, in the presence of the
non-minimal coupling between induced gravity and the scalar field,
both non-minimal coupling and dissipation affect dynamics of these
perturbations in relatively complicated manner. In the minimal case
and within the standard model, if we consider a small dissipation by
setting  $\Gamma\simeq 0$, the entropy perturbation vanishes for
long wavelength and the primordial spectrum of perturbation is due
to adiabatic perturbations [32]. But in our case, if we set
$\Gamma\simeq 0$, for long wavelength that $k\simeq 0$, the entropy
perturbation is given by
$$\delta P_{nad}=-2\mu^{2}(H\phi+\dot{\psi})\chi +\mu^{2}\delta
E_{\,0}^{0}{\cal{Y}}+\frac{2}{3}\mu^{2}\delta
E_{\,0}^{0}\big({\cal{Z}}-1\big)$$
\begin{equation}
+\Big(\delta
P_{\xi}-\delta\rho_{\xi}-2\frac{V_{,\varphi}}{\dot{\varphi}}
\int(\delta T_{i}^{0})_{\xi}\,dx_{i}+\frac{\rho_{0}V_{,\varphi}
\eta}{\rho_{\varphi}\dot{\varphi}}\int\delta E_{\,i}^{0}\,dx_{i}
\Big){\cal{Z}},
\end{equation}
where
$$\chi=\frac{\Bigg[\dot{\rho_{\xi}}\Big(-\frac{V_{,\varphi}}{\dot{\varphi}}+
\frac{\xi
f(R)\varphi}{\dot{\varphi}}\Big)+\dot{P_{\xi}}\Big(\frac{V_{,\varphi}}{\dot{\varphi}}-
\frac{\xi f(R)\varphi}{\dot{\varphi}}\Big)+2 V_{,\varphi}\big(\xi
f(R)\varphi-V_{,\varphi}\big)
\Bigg]{\cal{Z}}}{\big(V_{,\varphi}\dot{\varphi}+\dot{\rho_{\xi}}\big){\cal{Z}}
+4\mu^{2}E_{\,0}^{0}H\big({\cal{Z}}-1\big)}$$
\begin{equation}
+2\frac{V_{,\varphi}\rho_{0}\eta}{\rho_{\varphi}\dot{\varphi
}}{\cal{Z}}-\frac{4\mu^{2}E_{\,0}^{0}\Big(4\big({\cal{Z}}-1\big)-3{\cal{Y}}\Big)H^{2}
}{\big(V_{,\varphi}\dot{\varphi}+\dot{\rho_{\xi}}\big){\cal{Z}}
+4\mu^{2}E_{\,0}^{0}H\big({\cal{Z}}-1\big)}.
\end{equation}
In the absence of the non-minimal coupling, the entropy perturbation
reduces to the result of the minimal setup [18]
$$\delta
P_{nad}=4\mu^{2}(H\phi+\dot{\psi})
\Bigg[\frac{V_{,\varphi}^{2}{\cal{Z}}}{V_{,\varphi}\dot{\varphi}{\cal{Z}}
+4\mu^{2}E_{\,0}^{0}H\big({\cal{Z}}-1\big)}-\frac{V_{,\varphi}\rho_{0}\eta}{\rho_{\varphi}\dot{\varphi
}}{\cal{Z}}+\frac{2\mu^{2}E_{\,0}^{0}\Big(4\big({\cal{Z}}-1\big)-3{\cal{Y}}\Big)H^{2}
}{V_{,\varphi}\dot{\varphi}{\cal{Z}}
+4\mu^{2}E_{\,0}^{0}H\big({\cal{Z}}-1\big)}\Bigg]$$
$$+\mu^{2}\delta
E_{\,0}^{0}{\cal{Y}}+\frac{2}{3}\mu^{2}\delta
E_{\,0}^{0}\big({\cal{Z}}-1\big)+\frac{\rho_{0}V_{,\varphi}
\eta}{\rho_{\varphi}\dot{\varphi}}\int\delta E_{\,i}^{0}\,dx_{i}
{\cal{Z}}.$$ In the minimal standard theory of cosmological
perturbations, when there are no dissipation effects, all
perturbations are adiabatic and there is no trace of the
non-adiabatic perturbations. But, as we have shown here, in a
DGP-inspired non-minimal setup, in the absence of dissipations there
is a non-vanishing contribution of the non-adiabatic perturbations.
We note that our inspection shows that this effect is mainly as a
result of DGP than non-minimal coupling. The effect of the
non-minimal coupling tends to increase the
contribution of the entropy perturbations.\\
The curvature perturbation on a uniform density hypersurface is
defined as [42,43]
\begin{equation}
\zeta\equiv\psi+\frac{1}{6\mu^{2}}\frac{\delta\rho_{tot}}{\dot{H}}
\end{equation}
Using the acceleration equation
\begin{equation}
\dot{H}=-\frac{1}{2\mu^{2}}(\rho_{tot}+P_{tot})
\end{equation}
where by definition $\rho_{tot}=\rho_{eff}+\rho_{\gamma}$\,, the
curvature perturbation can be written as
\begin{equation}
\zeta\equiv\psi-\frac{\delta\rho_{tot}}{3\big(\rho_{tot}+P_{tot}\big)}.
\end{equation}
From this equation, we deduce [44]
\begin{equation}
\dot{\zeta}=H\Big(\frac{\delta P_{nad}}{\rho+P}\Big)_{tot},
\end{equation}
which implies that $\zeta$ is a constant if the pressure
perturbation is adiabatic on the large scales. This equation relates
the change in the comoving curvature perturbation due to the source
$\dot{P}\delta S$\,( or equivalently $ \delta P_{nad})$. Using (55)
for long wavelength perturbations, $\dot{\zeta}$ is given by
$$\dot{\zeta}=-\frac{2(H\phi+\dot{\psi})}{3H(1+\omega_{tot})}\chi+\frac{1}{3\mu^{2}H(1+\omega_{tot})}\Big(\delta
P_{\xi}-\delta\rho_{\xi}-2\frac{V_{,\varphi}}{\dot{\varphi}}
\int(\delta T_{i}^{0})_{\xi}\,dx_{i}+\frac{\rho_{0}V_{,\varphi}
\eta}{\rho_{\varphi}\dot{\varphi}}\int\delta E_{\,i}^{0}\,dx_{i}
\Big){\cal{Z}} $$
\begin{equation}
-\frac{2\rho_{\gamma}}{9\mu^{2}H(1+\omega_{tot})}\Big(\frac{4\rho_{0}aV_{,\varphi}v\eta}{
k\rho_{\varphi}\dot{\varphi}}{\cal{Z}}+\frac{\delta\rho_{\gamma}}{\rho_{\gamma}}
\big(1-\frac{3}{2}{\cal{Y}}\big)\Big)+\frac{\delta
E_{\,0}^{0}}{3H(1+\omega_{tot})}\Big({\cal{Y}}+\frac{2}{3}\big({\cal{Z}}-1\big)\Big).
\end{equation}
where $\omega_{tot}=\frac{P_{tot}}{\rho_{tot}}$. In contrast to the
minimal standard case, the entropy perturbations depend not only on
the dissipation effects but also they depends on the non-minimal
coupling of the scalar field and induced gravity in the DGP setup.
In other words, even with small dissipation, the entropy
perturbations are important yet. It was expected a priori, based on
the standard picture, that in the absence of dissipation the
perturbation should be adiabatic since just one field is present.
However, in our non-minimal DGP-inspired model with modified induced
gravity the effects of the non-adiabatic perturbations are present
yet and in this case curvature perturbations cannot be constant in
time and they attain an explicit time-dependence. These are new
results for the rest of the theory of cosmology perturbations. We
note that isocurvature perturbations are free to evolve on
superhorizon scales, and the amplitude at the present day depends on
the details of the entire cosmological evolution from the time that
they are formed. On the other hand, because all super-Hubble radius
perturbations evolve in the same way, the shape of the isocurvature
perturbation spectrum is preserved during this evolution [45].

\section{ The Power Spectrum}
In the previous section, we have shown that in the warm inflationary
model the entropy perturbations are generated since inflaton and
radiation fields interact with each other. Here we are going to
obtain scalar and tensorial perturbation for warm inflation and we
expect that for $\Gamma=0$, the results of cool inflation will be
recovered.\\
We take into account the slow-roll approximation at the large
scales, $k\ll aH$, where we need to describe the non-decreasing
adiabatic and entropy modes. In this situation, equations (51) and
(52) become respectively
\begin{equation}
(3H+\Gamma)\delta\dot{\varphi}+\Big(V_{,\varphi\varphi}
+\Gamma_{,\varphi}\dot{\varphi}-\xi f(R)\Big)\delta\varphi
\simeq\phi\big(2\xi f(R)\varphi-2V_{,\varphi}+\Gamma\dot{\varphi}
\big)-12\xi f'(R)\varphi\phi(\dot{H}+2H^{2}),
\end{equation}
\begin{equation}
\frac{\delta\rho_{\gamma}}{\rho_{\gamma}}\simeq\frac{\Gamma_{,\varphi}}{\Gamma}\delta\varphi
-3\phi,
\end{equation}
and equation (34) takes the following form
\begin{equation}
\phi\simeq\frac{1}{2\mu^{2}H}\Big(\frac{\Gamma}{4H}+\frac{\Gamma_{,\varphi}\dot{\varphi}}{
48H^{2}}+\frac{\rho_{\varphi}}{\rho_{0}\eta}+\frac{\mu^{2}}{\dot{\varphi}\delta\varphi}\int
\delta E_{i}^{0}\,d
x_{i}\Big)\Big[1+\xi\frac{\rho_{\varphi}}{\rho_{0}
\eta\mu^{2}}\varphi^{2}f'(R)\Big]^{-1} \dot{\varphi}\delta\varphi,
\end{equation}
where we have used the relation of the velocity field as
$v\simeq-\frac{k}{4aH}
\Big(\phi+\frac{\delta\rho_{\gamma}}{4\rho_{\gamma}}+\frac{3\Gamma\dot{\varphi}}{4\rho
_{\gamma}}\delta\varphi\Big)$ and $\int(\delta T_{i}^{0})_{\xi}\,d
x_{i}=2\xi\varphi^{2}f'(R)(\dot{\psi}+H\phi)$. Now solve these three
equations to find the desired relations. First, by substituting
equation (69) into equation (67), we find
$$(3H+\Gamma)\delta\dot{\varphi}+\Big(V_{,\varphi\varphi}
+\Gamma_{,\varphi}\dot{\varphi}-\xi f(R)\Big)\delta\varphi
\simeq\frac{1}{2\mu^{2}H}\Big(2\xi f(R)\varphi-2V_{,\varphi}
+\Gamma\dot{\varphi}-12\xi f'(R)\varphi(\dot{H}+2H^{2})\Big)$$
\begin{equation}
\times\Big(\frac{\Gamma}{4H}+\frac{\Gamma_{,\varphi}\dot{\varphi}}{
48H^{2}}+\frac{\rho_{\varphi}}{\rho_{0}\eta}+\frac{\mu^{2}}{\dot{\varphi}\delta\varphi}\int
\delta E_{i}^{0}\,d
x_{i}\Big)\Big[1+\xi\frac{\rho_{\varphi}}{\rho_{0}
\eta\mu^{2}}\varphi^{2}f'(R)\Big]^{-1}\dot{\varphi}\delta\varphi .
\end{equation}
Following [45], we define an auxiliary function as
\begin{equation}
\chi=\frac{\delta\varphi}{V_{,\varphi}}exp\Big(\int\frac{\Gamma_{,\varphi}}{\Gamma+3H}\,d\varphi\Big).
\end{equation}
Therefore, equation (70) can be rewritten as
$$\frac{\chi_{,\varphi}}{\chi}\simeq-\frac{9}{8}\frac{(\Gamma+2H)}{(\Gamma+3H)^{2}}
\Big[V_{,\varphi}-\xi f(R)\varphi+4\frac{(\Gamma+3H
)}{(\Gamma+2H)}\xi f'(R)\varphi(\dot{H}+2H^{2})\Big]$$
$$\times\Bigg[\Gamma-\frac{\Gamma_{,\varphi}\big(V_{,\varphi}-\xi
f(R)\varphi\big)}
{12H(\Gamma+3H)}+4H\frac{\rho_{\varphi}}{\rho_{0}\eta}+\frac{4H\mu^{2}}{\dot{\varphi}\delta\varphi}\int
\delta E_{i}^{0}\,d x_{i}\Bigg]\Big[1+\xi\frac{
\rho_{\varphi}}{\rho_{0}\eta\mu^{2}}\varphi^{2}f'(R)\Big]^{-1}$$
\begin{equation}
\times\bigg[\rho_{\varphi}+\rho_{0}+\varepsilon\rho_{0}\Big[{\cal{A}}_{0}^{2}+\frac{2
\eta}{\rho_{0}}\big(\rho_{\varphi}-\mu^{2}E_{\,0}^{0}\big)\Big]^{1/2}\bigg]^{-1}.
\end{equation}
A solution of this equation is given as $\chi=C\,
\exp\big(\int\frac{\chi'}{\chi}d\varphi\big)$, where $C$ is an
integration constant. From equation (71), $\delta\varphi$ is given
by
$$\delta\varphi\simeq C\,V_{,\varphi}exp\Bigg[-\int\Bigg(\frac{\Gamma_{,\varphi}}{\Gamma+3H}+
\frac{9}{8}\frac{(\Gamma+2H)}{(\Gamma+3H)^{2}} \Big[V_{,\varphi}-\xi
f(R)\varphi+4\frac{(\Gamma+3H )}{(\Gamma+2H)}\xi
f'(R)\varphi(\dot{H}+2H^{2})\Big]$$
$$\times\Bigg[\Gamma-\frac{\Gamma_{,\varphi}\big(V_{,\varphi}-\xi
f(R)\varphi\big)}
{12H(\Gamma+3H)}+4H\frac{\rho_{\varphi}}{\rho_{0}\eta}+\frac{4H\mu^{2}}{\dot{\varphi}\delta\varphi}\int
\delta E_{i}^{0}\,d x_{i}\Bigg]\Big[1+\xi\frac{
\rho_{\varphi}}{\rho_{0}\eta\mu^{2}}\varphi^{2}f'(R)\Big]^{-1}$$
\begin{equation}
\times\bigg[\rho_{\varphi}+\rho_{0}+\varepsilon\rho_{0}\Big[{\cal{A}}_{0}^{2}+\frac{2
\eta}{\rho_{0}}\big(\rho_{\varphi}-\mu^{2}E_{\,0}^{0}\big)\Big]^{1/2}\bigg]^{-1}\Bigg)d
\varphi\Bigg].
\end{equation}
For simplicity we define the following quantity
$$\textbf{A}(\varphi)\equiv-\int\Bigg(\frac{\Gamma_{,\varphi}}{\Gamma+3H}+
\frac{9}{8}\frac{(\Gamma+2H)}{(\Gamma+3H)^{2}} \Big[V_{,\varphi}-\xi
f(R)\varphi+4\frac{(\Gamma+3H )}{(\Gamma+2H)}\xi
f'(R)\varphi(\dot{H}+2H^{2})\Big]$$
$$\times\Bigg[\Gamma-\frac{\Gamma_{,\varphi}\big(V_{,\varphi}-\xi
f(R)\varphi\big)}
{12H(\Gamma+3H)}+4H\frac{\rho_{\varphi}}{\rho_{0}\eta}+\frac{4H\mu^{2}}{\dot{\varphi}\delta\varphi}\int
\delta E_{i}^{0}\,d x_{i}\Bigg]\Big[1+
\xi\frac{\rho_{\varphi}}{\rho_{0}\eta\mu^{2}}\varphi^{2}f'(R)\Big]^{-1}$$
\begin{equation}
\times\bigg[\rho_{\varphi}+\rho_{0}+\varepsilon\rho_{0}\Big[{\cal{A}}_{0}^{2}+\frac{2
\eta}{\rho_{0}}\big(\rho_{\varphi}-\mu^{2}E_{\,0}^{0}\big)\Big]^{1/2}\bigg]^{-1}\Bigg)d
\varphi.
\end{equation}
With this definition, equation (73) can be rewritten as
$$\delta\varphi\simeq C\,V_{,\varphi} \exp[\textbf{A}(\varphi)],$$ and
therefore, the density perturbation is given by ( see [45])
\begin{equation}
\delta_{H}=\frac{16}{5}\pi\mu^{2}\frac{\exp[-\textbf{A}(\varphi)]}{V_{,\varphi}}\delta\varphi.
\end{equation}
We note that the main result here is the presence of a non-adiabatic
pressure contribution due to the entropy perturbation. This pressure
controls the evolution of the curvature perturbation on large scales
during inflation. In fact in the presence of entropy perturbations
the primordial curvature perturbation is not constant after horizon
crossing, so the relevant value to be compared with observations
should be evaluated at most at the end of inflation. In this
respect, this quantity should be evaluated at the end of inflation
[45]. For $\Gamma=0$, this relation reduces to the density
perturbation in a cool non-minimal inflation model in the framework
of DGP-inspired modified gravity. In the high dissipation regime,
$r\gg1$, the fluctuations in the warm inflationary model generate by
thermal interactions rather than quantum fluctuations [46]
\begin{equation}
\delta{\varphi}^{2}=\frac{K_{f}T}{2\pi^{2}},
\end{equation}
where the freeze-out scale at which dissipation damps out the
thermally excited fluctuations is defined as
$$K_{f}\equiv\sqrt{\Gamma H}=\sqrt{3r}H \geq H.$$
Although we have used the usual spectrum for the field perturbations
in warm inflation, but the presence of the non-minimal coupling is
hidden in the Hubble parameter $H$. Given that this spectrum have
been derived in a complete different set-up, to begin with without
coupling of the scalar field to the Ricci scalar, it may hold in the
scenario studied in this work if we use the form of $H$ dependent on
the non-minimal coupling. It is obvious that modified induced
gravity on the brane shows itself in the Friedmann equation and
hence $H$.

Now equations (74) and (75) for $r\gg1$ can be rewritten as follows
$$\widehat{\textbf{A}}(\varphi)\equiv-\int\Bigg(\frac{\Gamma_{,\varphi}}{3Hr}+
\frac{9}{8}\Big[V_{,\varphi}-\xi
\varphi\big(f(R)+4f'(R)(\dot{H}+2H^{2})\big)\Big]$$
$$\times\Bigg[1-\frac{\Gamma_{,\varphi}\big(V_{,\varphi}-\xi
f(R)\varphi\big)}
{12H(3Hr)^{2}}+\frac{4}{3r}\frac{\rho_{\varphi}}{\rho_{0}\eta}+\frac{4\mu^{2}}{3r
\dot{\varphi}\delta\varphi}\int\delta E_{i}^{0}\,d x_{i}
\Bigg]\Big[1+
\xi\frac{\rho_{\varphi}}{\rho_{0}\eta\mu^{2}}\varphi^{2}f'(R)\Big]^{-1}$$
\begin{equation}
\times\bigg[\rho_{\varphi}+\rho_{0}+\varepsilon\rho_{0}\Big[{\cal{A}}_{0}^{2}+\frac{2
\eta}{\rho_{0}}\big(\rho_{\varphi}-\mu^{2}E_{\,0}^{0}\big)\Big]^{1/2}\bigg]^{-1}\Bigg)d
\varphi,
\end{equation}
and
\begin{equation}
\delta_{H}^{2}=\frac{128}{25}\mu^{4}exp[-2\widehat{\textbf{A}}(\varphi)]\frac{H\sqrt{3r}T
}{V_{,\varphi}^{2}},
\end{equation}
where a hat on a quantity shows that quantity is computed in the
high dissipation regime. One important quantity in the inflationary
cosmology is the scalar spectral index defined as follows
\begin{equation}
n_{s}=1+\frac{d\ln\delta_{H}^{2}}{d\ln k}.
\end{equation}
In our setup, this quantity in the high-dissipation regime, $r\gg
1$, becomes
$$\widehat{n_{s}}=1-\widehat{\epsilon}+\frac{3}{4}\widehat{\alpha}-\frac{\mu^{2}}{r}
\Bigg[\frac{3}{4}\Bigg(V_{,\varphi\varphi}-\xi\Big(f'(R)\dot{R}+f(R)\Big)\Bigg)
+2\Big(\xi
f(R)\varphi-V_{,\varphi}\Big)\Big(\widehat{\textbf{A}}_{,\varphi}+\frac{V_{,\varphi
\varphi}}{V_{,\varphi}}\Big)\Bigg]$$
\begin{equation}
\times\bigg[\rho_{\varphi}+\rho_{0}+\varepsilon\rho_{0}\Big[{\cal{A}}_{0}^{2}+\frac{2
\eta}{\rho_{0}}\big(\rho_{\varphi}-\mu^{2}E_{\,0}^{0}\big)\Big]^{1/2}\bigg]^{-1},
\end{equation}
where $\widehat{\textbf{A}}_{,\varphi}$ is the integrand of equation
(77). The running of the spectral index in our setup is given as
follows
$$\frac{d\widehat{n_{s}}}{d\ln k}=\hspace{15cm}$$
$$=-2\widehat{\epsilon}\,^{2}+\widehat{\epsilon}\,
\widehat{\alpha}+\widehat{\epsilon}\mu^{2}\Big[-2\frac{\widehat{\textbf{A}}_{,\varphi}}
{r}+\frac{3r_{,\varphi}}{4r^{2}}-2\frac{V_{,\varphi\varphi}}{rV_{,\varphi}}\Big]
\Big(\xi f(R)\varphi-V_{,\varphi}\Big)\bigg[\rho_{\varphi}+\rho_{0}+
\varepsilon\rho_{0}\Big[{\cal{A}}_{0}^{2}+\frac{2
\eta}{\rho_{0}}\big(\rho_{\varphi}-\mu^{2}E_{\,0}^{0}\big)\Big]^{1/2}\bigg]^{-1}$$
$$+\mu^{4}\Bigg[\frac{
\widehat{\textbf{A}}_{,\varphi\varphi}}{r^{2}}+\frac{3r_{,\varphi\varphi}}{4r^{3}}
-\frac{3r_{,\varphi}^{2}}{4r^{4}}-2\frac{V_{,\varphi\varphi\varphi}}{r^{2}V_{,\varphi}}
+2\frac{V_{,\varphi\varphi}^{2}}{r^{2}V_{,\varphi}^{2}}\Bigg]\Big(\xi
f(R)\varphi-V_{,\varphi}\Big)^{2}$$
\begin{equation}
\times\bigg[\rho_{\varphi}+\rho_{0}+
\varepsilon\rho_{0}\Big[{\cal{A}}_{0}^{2}+\frac{2
\eta}{\rho_{0}}\big(\rho_{\varphi}-\mu^{2}E_{\,0}^{0}\big)\Big]^{1/2}\bigg]^{-2}.
\end{equation}
For an inflationary model driven by just one scalar field, the
running of the spectral index constraint by the WMAP5+SDSS+SNIa
combined data is $\alpha_{s}\equiv \frac{d\widehat{n_{s}}}{d\ln
k}=-0.032^{+0.021}_{-0.020}$, with $1\sigma$ CL \, ( see for
instance [47] and references therein). In our case, we see that
dissipative effects, modified induced gravity and the non-minimal
coupling of the scalar field and induced gravity have the potential
to produce a variety of spectra ranging between red and blue ( see
[3,6,8,17,32,40] for realization of these spectral index in
different scenarios).

Now we pay attention to the tensorial perturbations. As it has been
mentioned in Ref.[48], the generation of tensor perturbations during
inflation period produces stimulated emission in the thermal
background of gravitational waves. This process changes the power
spectrum of the tensor modes by an extra, temperature-dependent
factor given by $\coth(\frac{k}{2T})$. So, the spectrum of tensor
perturbations is given by
\begin{equation}
A_{g}^{2}=\frac{1}{6\pi^{2}\mu^{4}}\coth\big(\frac{k}{2T}\big)\bigg
[\rho_{\varphi}+\rho_{0}+\varepsilon\rho_{0}\Big[{\cal{A}}_{0}^{2}+\frac{2
\eta}{\rho_{0}}\big(\rho_{\varphi}-\mu^{2}E_{\,0}^{0}\big)\Big]^{1/2}\bigg].
\end{equation}
Using equations (78) and (82), in the limit of $r\gg 1$ the tensor
to scalar ratio is given by
\begin{equation}
{\cal{R}}=\Big(\frac{A_{g}^{2}}{\cal{P}_{R}}\Big)_{k=k_{i}}\simeq
\frac{1}{64\sqrt{3}\pi^{2}\mu^{6}}\Bigg[\frac{V_{,\varphi}^{2}\,H}
{T\,\sqrt{r}}\,exp\,[2\widehat{\textbf{A}}(\varphi)]\coth\big(\frac{k}{2T}\big)\Bigg]_{k=k_{i}},
\end{equation}
where ${\cal{P}_{R}}=\frac{25}{4}\delta_{H}^{2}$ and $k_{i}$ denotes
the value of $k$ when universe scale crosses the Hubble horizon
during inflation. The WMAP5+SDSS+SNIa combined data gives the values
of the scalar curvature spectrum as
${\cal{P}_{R}}(k_{i})\equiv\frac{25}{4}\delta_{H}^{2}=
(2.445\pm0.096)\times 10^{-9}$\, at $k_{i}=0.002 Mpc^{-1}$\, and the
tensor to scalar ratio at this value of $k_{i}$ as
${\cal{R}}(k_{i})<0.22$ \, [49]. Evidently, these values will set
severe constraints on the parameters of our model some of which are
studied in the next section.

\section{Numerics of the parameter space}
Now we study numerically the case with the following scalar field
potential
\begin{equation}
V(\varphi)=V_{0}\exp\Big(-\sqrt{\frac{2}{p}}\frac{\varphi}{\mu}\Big),
\end{equation}
where $V_{0}$ and $p$ are constants. We consider a modified gravity
model with\,  $f(R)=f_{0}R^{n}$,\, where $f_{0}$ and $n$ are
constant [50]. We set also\,
$\Gamma(\varphi)\equiv\big[\upsilon+1+\varepsilon
({\cal{A}}_{0}^{2}+2\eta\upsilon)^{1/2}\big]^{\frac{1}{2}}$,\,( see
[18]), and we will restrict ourselves to the high dissipation regime
where $r\gg 1$. In our presentation of the numerical results we take
a dissipative coefficient proportional to the Hubble parameter, so
that $r$ is constant. In our calculations we use $r=10000$ since we
consider the high dissipation regime, $r\gg 1$. As we will show, the
value of the dissipation coefficient has some impact on the results
as it is usually the case in the standard warm inflation. We will
come back to the role played by dissipation shortly.

From equation (78), the scalar power spectrum in our model with
exponential potential (84) is given by
\begin{equation}
{\cal{P}_{R}}(k_{i})=\frac{16\mu^{\frac{11}{2}}}{3^{\frac{1}{4}}}\Bigg[\Big(\frac{pT}{\upsilon^{
\frac{7}{4}}\rho_{0}^{2}}\Big)\exp[-2\widehat{\textbf{A}}(\varphi)]\big[x+1+\varepsilon
\omega_{x}\big]^{\frac{1}{4}}\Big[1+\frac{1}{\upsilon}(1+\varepsilon\omega)\Big]^
{\frac{1}{4}}\Bigg]_{k=k_{i}},
\end{equation}
where by definition
$$\widehat{\textbf{A}}(\varphi)=\int\Bigg(\frac{1}{2\mu}\sqrt{\frac{2}{p}}
\Big[1+\frac{1}{\upsilon}(1+\varepsilon\omega)\Big]^{-1}\big(1+\varepsilon
\eta\omega^{-1}\big)+\frac{9}{8}\Big[\sqrt{\frac{2}{p}}\frac{\rho_{0}\upsilon}
{\mu}+\xi\varphi\Big(f_{0}\big(1+\frac{2}{3}n\big)R^{n}\Big)\Big]$$
$$\times\Bigg[1-\frac{1}{24\upsilon^{\frac{1}{2}}}\sqrt{\frac{6}{p}}
\Big[1+\frac{1}{\upsilon}(1+\varepsilon\omega)\Big]^{-\frac{3}{2}}\big[
1+\varepsilon\eta\omega^{-1}\big]\Big(\sqrt{\frac{2}{p}}\frac{\rho_{0}
\upsilon}{\mu}+\xi f_{0}R^{n}\varphi\Big)\big[x+1+\varepsilon
\omega_{x}\big]^{-\frac{1}{2}}$$
$$+\frac{4}{\sqrt{3}\mu\upsilon^{\frac{1}{2}}}\Big(\frac{x}{\eta}+\frac{\mu^{2}}{
\dot{\varphi}\delta\varphi}\int\delta E_{i}^{0}\,d x_{i}
\Big)\big[x+1+\varepsilon
\omega_{x}\big]^{\frac{1}{2}}\Big[1+\frac{1}{\upsilon}(1+\varepsilon\omega)\Big]^
{-\frac{1}{2}}\Bigg]$$
\begin{equation}
\times\Big[1+\xi\frac{x\varphi^{2}}{\eta\mu^{2}} n
f_{0}R^{n-1}\Big]^{-1}\big[x+1+\varepsilon
\omega_{x}\big]^{-1}\Bigg)d\varphi,
\end{equation}
and other quantities are defined as follows
\begin{equation}
\upsilon\equiv\frac{V(\varphi)}{\rho_{0}},\,\,\,\,\,\omega\equiv\Big
({\cal{A}}_{0}^{2}+2\eta\upsilon\Big)^{1/2},\,\,\,\,\,\,x\equiv\frac{1}{\rho_{0}}
\big(\rho_{\varphi}-\mu^{2}E_{\,0}^{0}\big),\,\,\,\,\,\omega_{x}\equiv\Big
({\cal{A}}_{0}^{2}+2\eta x\Big)^{1/2}.
\end{equation}
From equation (83), the tensor to scalar ratio in this setup is
given by
\begin{equation}
{\cal{R}}(k_{i})\simeq\frac{1}{32\times
3^{\frac{3}{4}}\pi^{2}\mu^{\frac{19}{2}}}\Bigg[\Big(\frac{\upsilon^
{\frac{7}{4}}\rho_{0}^{2}}{pT}\Big)\exp[2\widehat{\textbf{A}}(\varphi)]\big[x+1+\varepsilon
\omega_{x}\big]^{\frac{3}{4}}\Big[1+\frac{1}{\upsilon}
(1+\varepsilon\omega)\Big]^{-\frac{1}{4}}\coth\big(\frac{k}{2T}\big)\Bigg]_{k=k_{i}}.
\end{equation}
Equations (85) and (88) with the definitions (86) and (87) are very
complicated and to have an intuition, we have to study these
quantities numerically. Using the appropriate values of
${\cal{P}_{R}}(k_{i})$ and ${\cal{R}}(k_{i})$ as mentioned
previously, equations (85) and (88) lead us to the following result
\begin{equation}
3.056\times10^{-8}=\frac{x_{i}}{\mu^{4}}\coth\big(\frac{k_{i}}{2T}\big)
\Big[1+\frac{1}{x_{i}}+\frac{\varepsilon}{x_{i}}\omega_{x}\Big].
\end{equation}
Here the subscript $i$ means that the corresponding quantity should
be calculated at $k_{i}= aH$ where we set $k_{i}=0.002 Mpc^{-1}$.
From equation (89) we get
\begin{equation}
x_{i}=\frac{2(-1+D+\eta)}{(D-1)^{2}},
\end{equation}
where
\begin{equation}
D=\frac{3.056\times10^{-8}}
{x_{\mu}\coth\big(\frac{k_{i}}{2T}\big)}=\frac{3.056\times10^{-8}}
{\big(\upsilon_{\mu}+\frac{\rho^{(curve)}}
{\rho_{0}\mu^{4}}\big)\coth\big(\frac{k_{i}}{2T}\big)},
\end{equation}
and $x_{\mu}=\frac{x_{i}}{\mu^{4}}$,\,\,$\upsilon_{\mu}=\frac{
\upsilon_{i}}{\mu^{4}}$. In this analysis, we set ${\cal{A}}_{0}=1$
for both DGP-branches of the model. Also we set $p=50$,\,
$\eta=0.99$,\, $k_{i}=0.002 Mpc^{-1}$,\, $T=0.24\times10^{16}Gev$
and $\mu^{2}\sim(10^{17}Gev)^{2}$. Note that important quantities
such as $\frac{\lambda}{\mu^{4}}$\, and \, $\frac{m_{5}}{\mu}$ now
depend on the parameters of the model in this warped DGP-inspired
framework. The results of our numerical calculations are shown in
figures $1$, $2$, $3$ and $4$. Figure $1$ shows the spectral index
versus $x\equiv \frac{\rho_{\varphi}}{\rho_{0}}$ for normal (
$\varepsilon=-1$) branch of the model. Depending on the values of
the conformal coupling, it is possible to have both red and blue
spectrum in this model. We note that positive values of the
non-minimal coupling give more reliable results in comparison with
observations ( this is supported from other viewpoints too; see for
instance [27]). With positive $\xi$, our model favors only the red
power spectrum. Figure $2$ shows the running of the spectral index
in normal branch of the model. For $\xi=-\frac{1}{12}$, the
calculated running in our model cannot be compared with observations
and therefore is excluded from our consideration. Figures $3$ and
$4$ show the corresponding results for the self-accelerating branch
of the scenario. Again, negative values of the non-minimal coupling
are excluded on observational grounds. These values are essentially
related to anti-gravitation.

\begin{figure}[htp]
\begin{center}\includegraphics{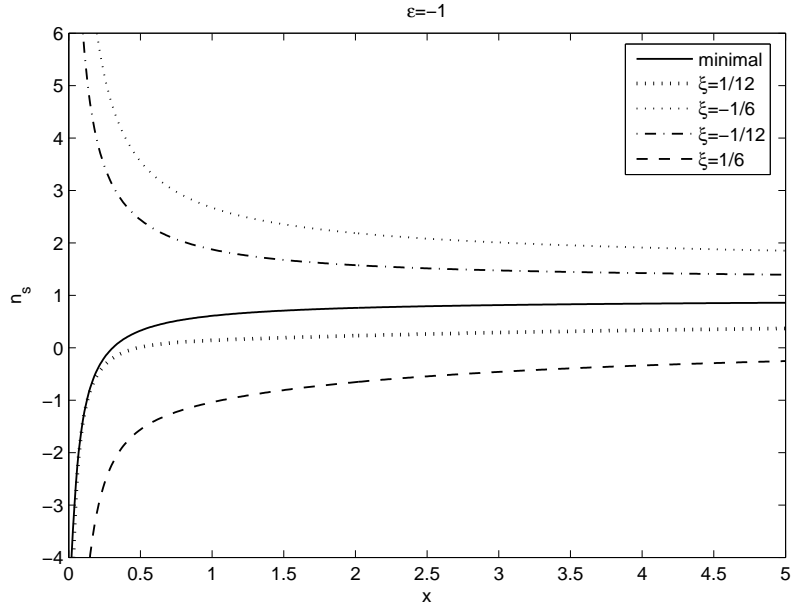} \vspace{8cm}
\end{center}
 \caption{\small { The spectral index versus $x\equiv
\frac{\rho_{\varphi}}{\rho_{0}}$ for normal ( $\varepsilon=-1$)
branch of the model.}}
\end{figure}

\begin{figure}[htp]
\begin{center}\includegraphics{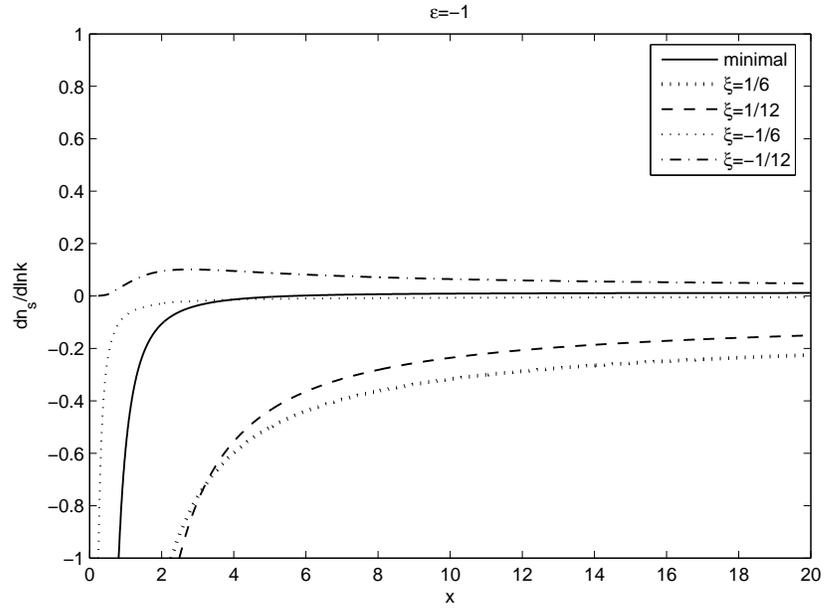} \vspace{6cm}
\end{center}
 \caption{\small {The running of the spectral index versus $x$ for
  normal branch of the model.}}
\end{figure}

\begin{figure}[htp]
\begin{center}\includegraphics{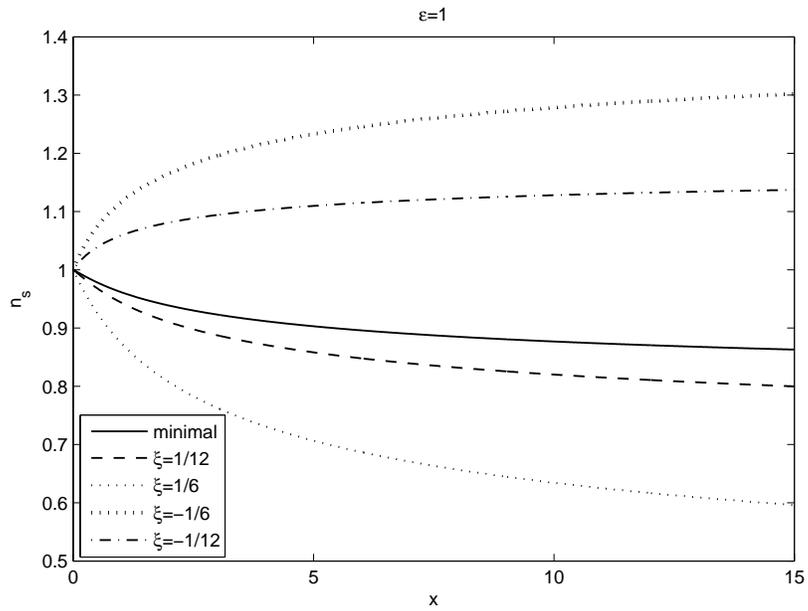} \vspace{6cm}
\end{center}
 \caption{\small {The spectral index versus $x\equiv
\frac{\rho_{\varphi}}{\rho_{0}}$ for self-accelerating (
$\varepsilon=+1$) branch of the model.}}
\end{figure}

\begin{figure}[htp]
\begin{center}\includegraphics{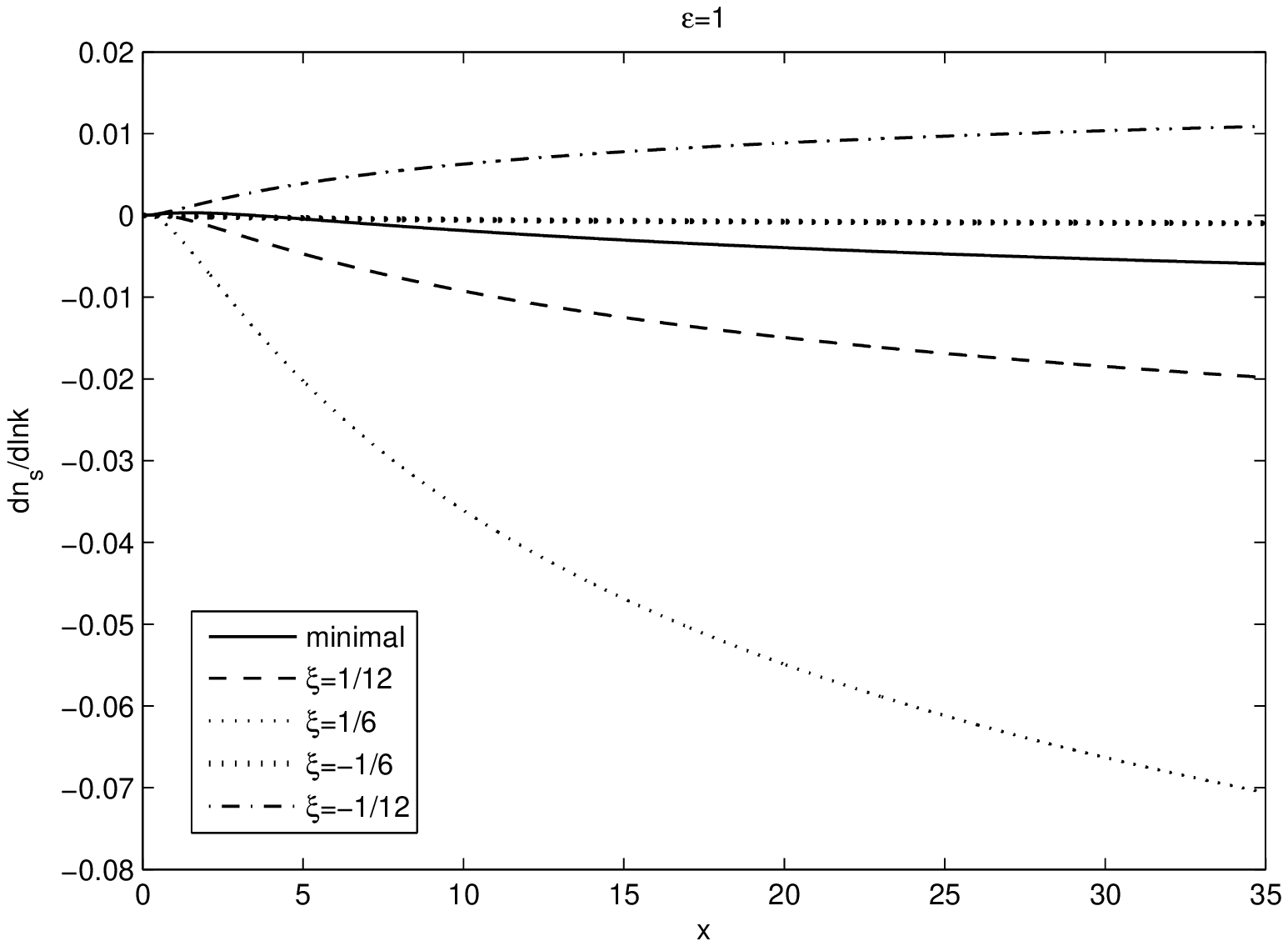} \vspace{5.3cm}
\end{center}
 \caption{\small {The running of the spectral index versus $x$ for self-accelerating (
$\varepsilon=+1$) branch of the model.}}
\end{figure}

Now we focus on the effect of dissipation on the inflation
parameters. Figure 5 shows the variation of the slow-roll parameter
$\epsilon$ versus $x$ for different values of the dissipation
factor, $r$. As this figure shows, $\epsilon$ decreases by
increasing the dissipation effect for a fixed value of $x$.

\begin{figure}[htp]
\begin{center}\includegraphics{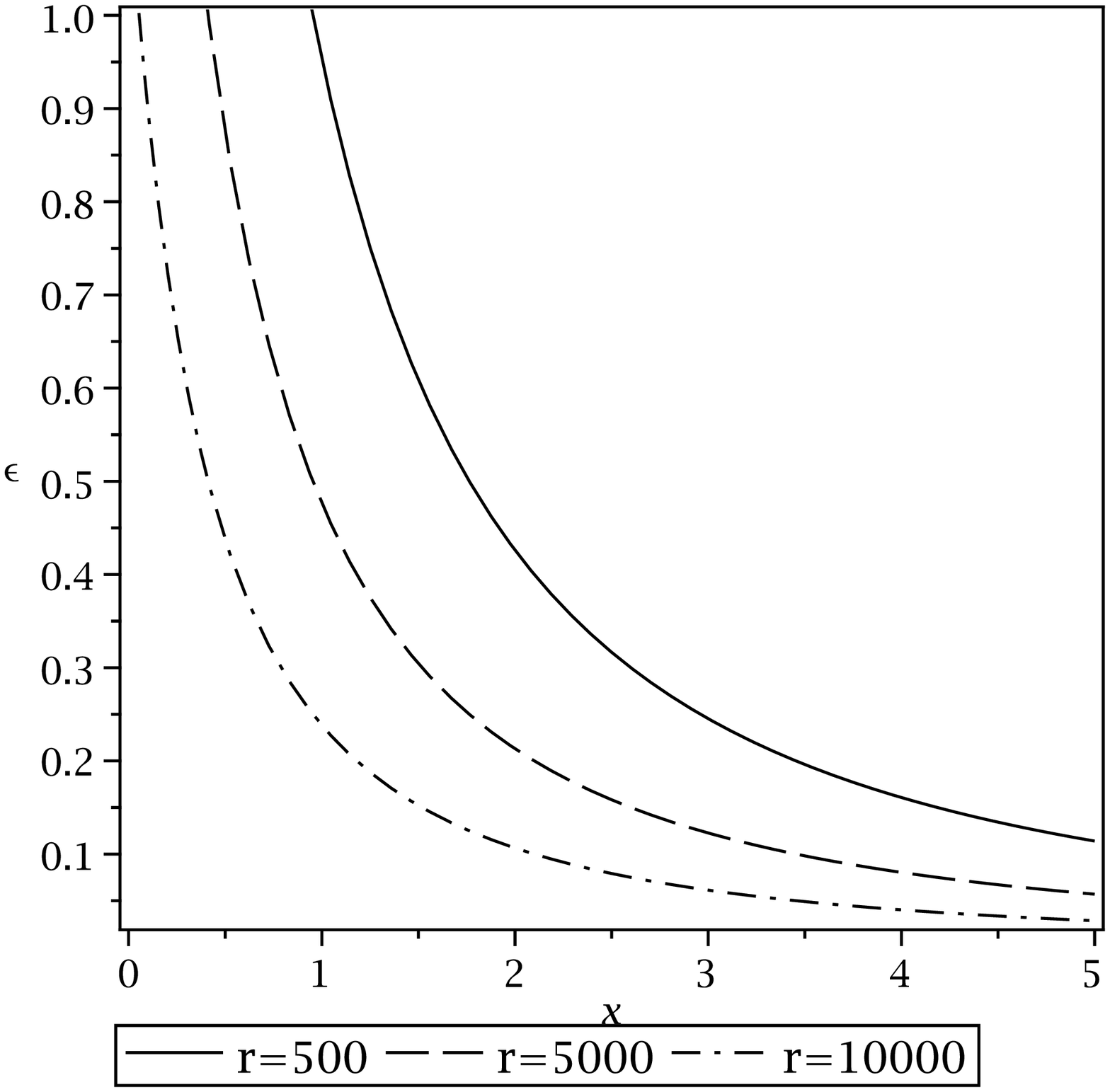} \vspace{8cm}
\end{center}
 \caption{\small { The slow-roll parameter $\epsilon$ versus $x\equiv
\frac{\rho_{\varphi}}{\rho_{0}}$ for self-accelerating (
$\varepsilon=+1$) branch of the model for different values of the
dissipation factor, $r$. }}
\end{figure}

In figure 6 we have considered the situation for a continuous
variation of the dissipation parameter. Again this figure shows that
the slow-roll parameter decreases by increasing $r$. However,
$\epsilon$ increases by increasing $x$ where $x\equiv
\frac{\rho_{\varphi}}{\rho_{0}}$.

\begin{figure}[htp]
\begin{center}\includegraphics{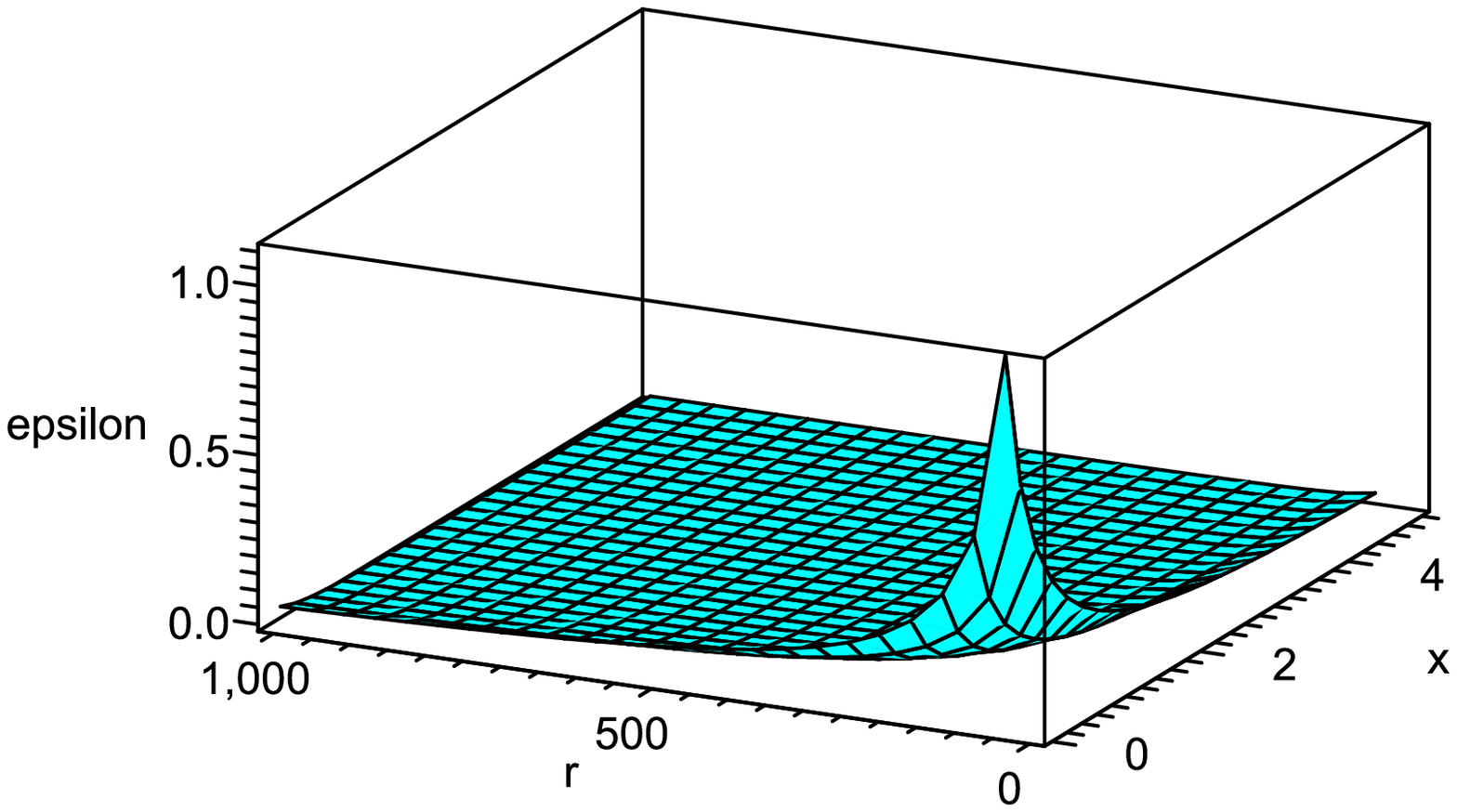} \vspace{5cm}
\end{center}
 \caption{\small { The slow-roll parameter $\epsilon$ versus $x\equiv
\frac{\rho_{\varphi}}{\rho_{0}}$ and $r$ for self-accelerating (
$\varepsilon=+1$) branch of the model. }}
\end{figure}

Now to see the effect of dissipation on the number of e-folds, we
define
\begin{equation}
g\equiv-\frac{1}{\mu^2}\frac{(1+r)}{\Big(V_{,\varphi}-\xi
f(R)\varphi\Big)}\bigg[\rho_{\varphi}+\rho_{0}+\varepsilon
\rho_{0}\Big({\cal{A}}_{0}^{2}+\frac{2\eta\rho_{\varphi}}{\rho_{0}}\Big)^{1/2}\bigg]\,,
\end{equation}
which is the integrand of the number of e-folds defined as (27). We
rephrase this quantity and also equation (27) versus $x$. Figure 7
shows the variation of $g$ versus $x$ for different values of the
dissipation factor, $r$. The number of e-folds is given by the
surface enclosed between this curve and two vertical lines located
at $x_{i}$ and $x_{e}$ corresponding to $\varphi_{i}$ and
$\varphi_{e}$ respectively. As this figure shows, the number of
e-folds increases by increasing dissipation factor.

\begin{figure}[htp]
\begin{center}\includegraphics{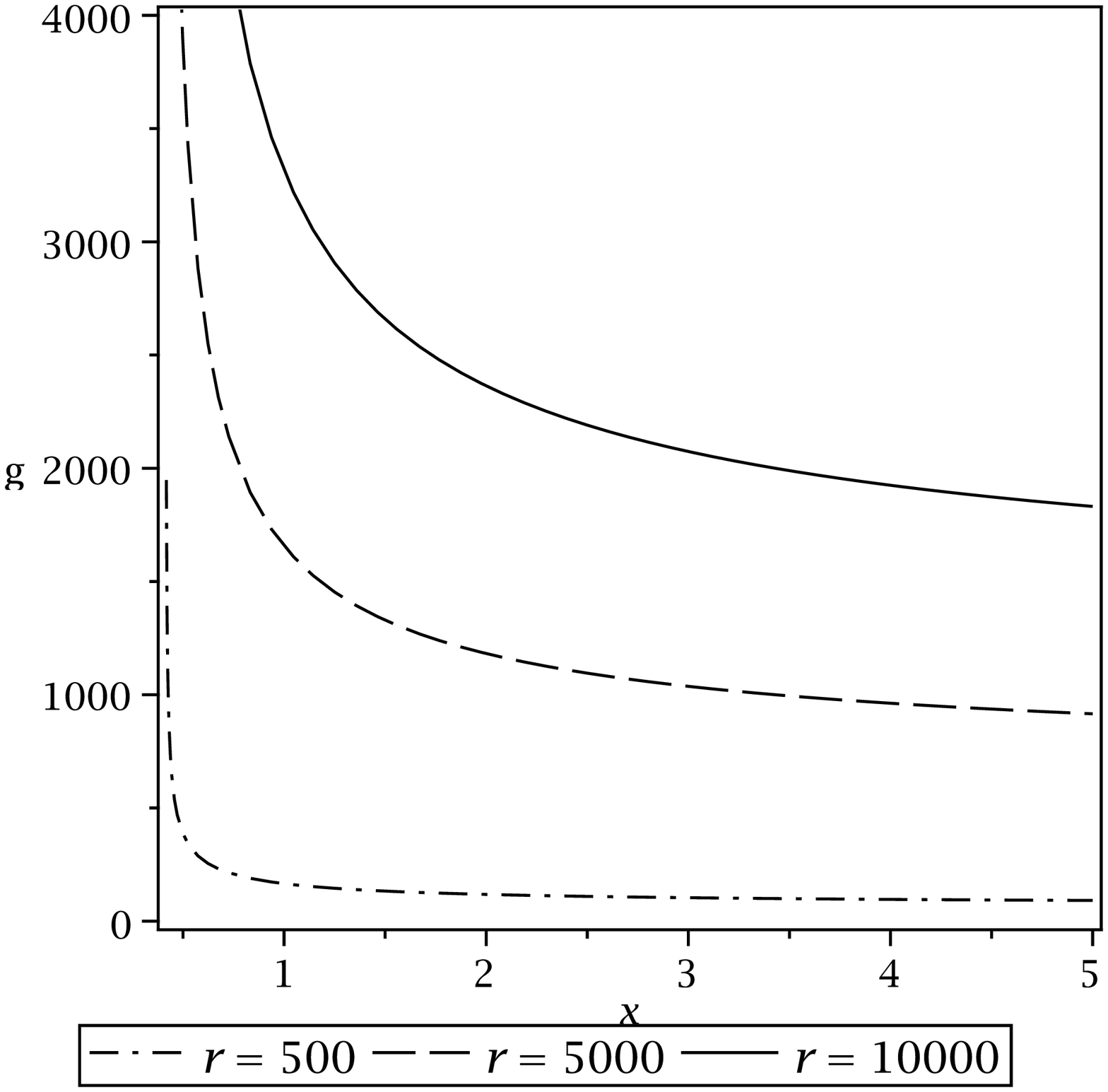} \vspace{9cm}
\end{center}
 \caption{\small { Variation of $g$ versus $x\equiv
\frac{\rho_{\varphi}}{\rho_{0}}$ for self-accelerating (
$\varepsilon=+1$) branch of the model for different values of the
dissipation factor, $r$. }}
\end{figure}

Finally, we note that the self-accelerating solution of the DGP
setup is unstable due to existence of ghosts [51,52]. In our setup,
incorporation of several new degrees of freedom such as modified
induced gravity, non-minimal coupling and the warped geometry of the
bulk has provided a relatively wider parameter space. This wider
parameter space may provide a better framework to treat
instabilities of the self-accelerating solution. However this is not
an easy task and lies out of our interest here ( for a recent
progress in this direction see [53]).

\section{Summary and Conclusions}
In this paper we have studied cosmological perturbations and their
evolution in a braneworld viewpoint of the warm inflation in the
presence of interaction between inflaton and induced gravity on the
brane. We have incorporated possible modification of the induced
gravity on the brane in the spirit of $f(R)$-gravity. The
cosmological perturbations are treated with complete details and the
roles played by modification of the induced gravity, dissipation and
the non-minimal coupling are discussed. The main results of our
study can be summarized as follows: in the minimal standard theory
of cosmological perturbations, when there are no dissipation
effects, all perturbations are adiabatic and there is no trace of
the non-adiabatic perturbations. But, as we have shown here, in a
DGP-inspired non-minimal setup with modified induced gravity, in the
absence of dissipation there is a non-vanishing contribution of the
non-adiabatic perturbations. In this setup, this effect is mainly as
a result of DGP ( with modified induced gravity) than the
non-minimal coupling. The effect of the non-minimal coupling tends
to increase the contribution of the entropy perturbations. The
numerical analysis of the parameter space ( which is wide enough due
to incorporation of several new degrees of freedom) shows that it is
possible to have both red and blue spectrum and relatively large
running of the spectral index depending on the sign of the
non-minimal coupling. However, negative values of the non-minimal
coupling give results that are not supported by observations and
therefore should be excluded from our considerations. The effect of
dissipation is so that the slow-roll parameter $\epsilon$ decreases
by increasing the dissipation factor. However, the number of e-folds
increases by increasing the dissipation factor. Modification of the
induced gravity and its coupling to the inflaton field in a warm
inflation framework brings a variety of new possibilities which
constraint our model in comparison with the recent observations. In
other words, this wider parameter space provides much more freedom
than single, self-interacting scalar field inflation to fit with the
observational data.

\section{Acknowledgments}
We are indebted to an anonymous referee for insightful suggestions
and comments. Part of this work has been done during B. F.
sabbatical leave at the INFN Institute, Frascati, Italy. She would
like to express her deepest appreciation to the members of this
Institute, especially \emph{Professor Stefano Bellucci} for kind
hospitality and supports. The work of K. N. is supported financially
by the Research Institute for Astronomy and Astrophysics of Maragha,
Iran.

\end{document}